\documentclass[10pt,letterpaper]{article}

\usepackage{changepage}

\usepackage[utf8]{inputenc}

\usepackage{textcomp,marvosym}

\usepackage{fixltx2e}

\usepackage{amsmath,amssymb}

\usepackage{cite}

\usepackage{nameref,hyperref}

\usepackage[right]{lineno}

\usepackage{microtype}
\DisableLigatures[f]{encoding = *, family = * }

\usepackage{rotating}

\usepackage[usenames]{color}
\usepackage{graphics}
\usepackage{graphicx}
\usepackage{multimedia}
\usepackage{ulem}

\raggedright
\usepackage[aboveskip=1pt,labelfont=bf,labelsep=period,justification=raggedright,singlelinecheck=off]{caption}

\makeatletter
\renewcommand{\@biblabel}[1]{\quad#1.}
\makeatother

\date{}
\usepackage{lastpage,fancyhdr,graphicx}
\usepackage{epstopdf}
\fancyfootoffset[L]{2.25in}

\newcommand{\beginsupplement}{%
        \setcounter{table}{0}
        \renewcommand{\thetable}{S\arabic{table}}%
        \setcounter{figure}{0}
        \renewcommand{\thefigure}{S\arabic{figure}}%
     }

\everymath{\displaystyle}     

\begin{document}

\vspace*{0.35in}

\begin{flushleft}
{\Large
\textbf{Keratin Dynamics: Modeling the Interplay between Turnover and Transport}
}
\newline
\\
St\'ephanie Portet\textsuperscript{1,*},
Anotida Madzvamuse\textsuperscript{2},
Andy Chung\textsuperscript{2}, 
Rudolf E. Leube\textsuperscript{3}, 
Reinhard Windoffer\textsuperscript{3} 
\\
\bigskip
\bf{1} Department of Mathematics, University of Manitoba, Winnipeg, Manitoba, Canada
\\
\bf{2} Department of Mathematics, School of Mathematical and Physical Sciences, University of Sussex, Brighton, United Kingdom
\\
\bf{3} Institute of Molecular and Cellular Anatomy, Rheinisch-Westf\"alische Technische Hochschule Aachen University, Aachen, Germany
\\
\bigskip

* portets@cc.umanitoba.ca

\end{flushleft}

\section*{Abstract}
Keratin are among the most abundant proteins in epithelial cells. Functions of the keratin network in cells are shaped by their dynamical organization. Using a collection of experimentally-driven mathematical models, different hypotheses  for the turnover and transport of the keratin material in epithelial cells are tested. The interplay between turnover and transport and their effects on the keratin organization in cells are hence investigated by combining mathematical modeling and experimental data. Amongst the collection of mathematical models considered, a best model strongly supported by experimental data is identified.  Fundamental to this approach is the fact that optimal parameter values associated with the best fit for each model are established.  The best candidate among the best fits is characterized by the disassembly of the assembled keratin material in the perinuclear region and  an active transport of the assembled keratin. Our study shows that an active transport of the assembled keratin is required to explain the experimentally observed keratin organization.


\section*{Introduction}
\begin{figure}[!h]
		\includegraphics[height=7.0cm, width=8.0cm]{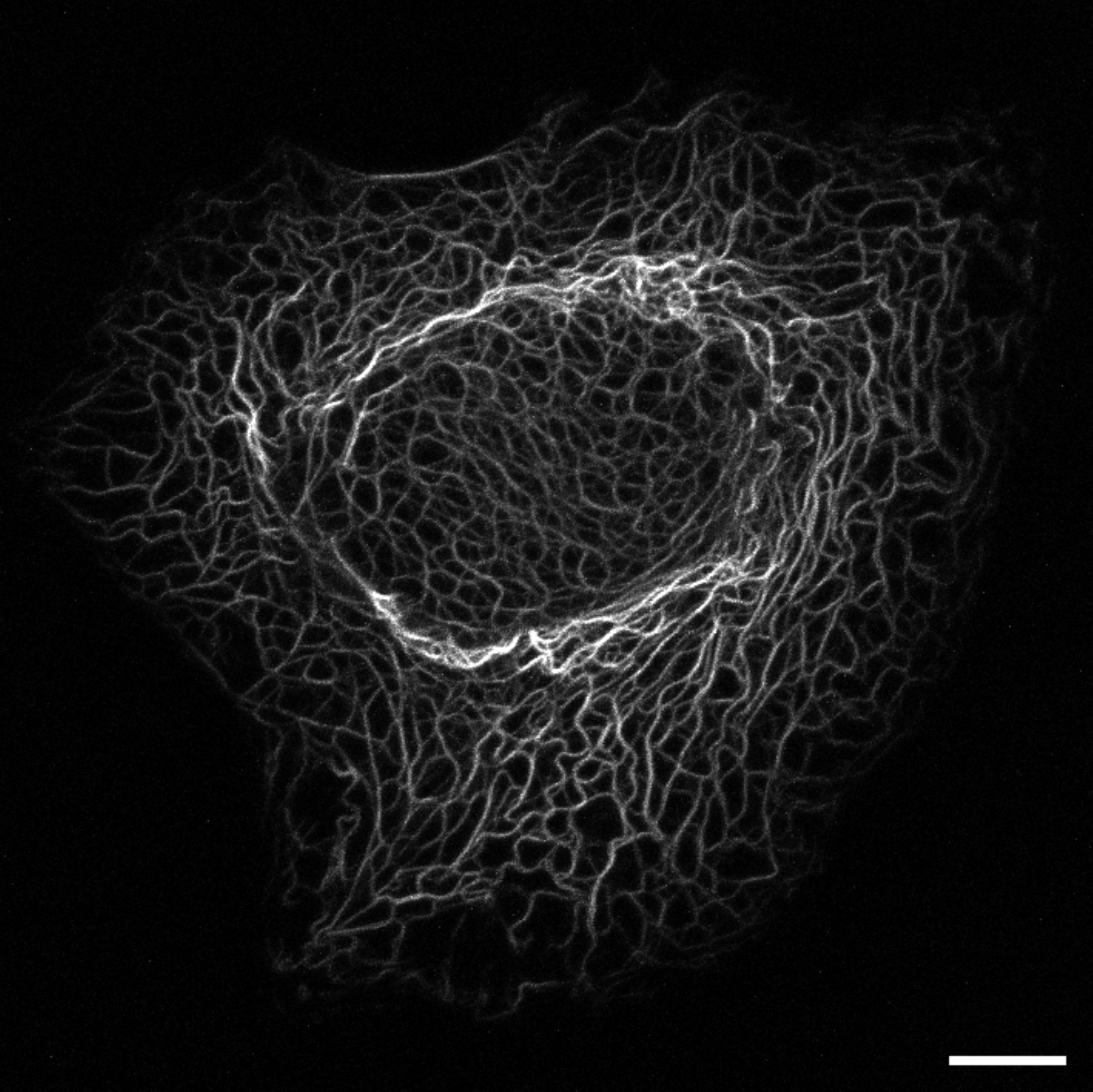}
\caption{ {\bf Keratin network.} Image taken from a time-lapse fluorescence recording  of a single hepatocellular carcinoma-derived PLC cell of clone PK18-5 \cite{Strnad2002} stably expressing fluorescent fusion protein HK18-YFP consisting of human keratin 18 and enhanced yellow fluorescent protein. Bar 10 $\mu m$.}
\label{fig:Keratin}
\end{figure}

The epithelial cytoskeleton is characterized by abundant keratin intermediate filaments (Fig.~\ref{fig:Keratin}). The cytoplasmic keratin filament network is responsible for the mechanical stress resistance of epithelial cells and contributes significantly to epithelial stiffness \cite{Ramms2013,Seltmann2013}.  The importance of keratins for epithelial tissue stability is reflected by a large group of genetic skin blistering diseases that are caused by point mutations in keratin-encoding genes \cite{Coulombe2009,Homberg2014}. The mechanical functions not only rely on static resilience but also necessitate a high degree of plasticity, for example in migrating cells during wound healing \cite{Coulombe2013}.  The current view is that keratins act as general stress absorbers protecting epithelial cells not only against mechanical insults but also against irradiation or osmotic and microbial challenges. Thus, keratins are involved in heat shock response, apoptosis and organelle homeostasis \cite{Toivola2010}. Furthermore, functions affecting processes such as proliferation, differentiation and inflammation are also dependent on keratins (see recent reviews in \cite{Pan2013,Homberg2014}).
\vspace{.5cm}

All of these functions are tightly coupled to the keratin network dynamics (see  \nameref{SI0}). Examination of cultured epithelial cells producing fluorescent keratins has provided evidence for the different mechanisms that are involved in the continuous renewal and reshaping of the keratin system \cite{Kolsch2010}. On the basis of these observations, a biological model of the keratin cycle has been proposed by Leube and Windoffer {\it et al.} in \cite{Leube2011,Windoffer2011}. This biological model takes into account the  assembly/disassembly and transport of keratins. In this study, it was proposed that assembly and disassembly occur in topologically defined regions with assembly taking place predominantly in the cell periphery while disassembly takes place primarily in the perinuclear region. The biological model further postulates active transport of insoluble assembly stages of keratins toward the nucleus and rapid diffusion of soluble subunits throughout the cytoplasm. To the best of our  knowledge, how these different processes are coupled and regulated is not yet known.

In a previous work we developed methods to examine and quantify the keratin transport and turnover in epithelial cells \cite{Moch2013}; the spatial distribution of the assembled keratin material in epithelial cells was available at 24 hours and 48 hours after seeding.  The current effort is to use the available qualitative and quantitative observations to derive, from first principles, experimentally-driven mathematical models that could yield hypothetical predictions testable in laboratories. Our approach is unique, it translates experimental observations and data into a series of alternative plausible mathematical models or scenarios  to further advance our understanding of the critical parameters in keratin cycling. Fundamental to this approach is the fact that optimal parameter values for each scenario are established and out of this set, a single scenario is identified that best fits experimental observations and data.

Hence, a collection of mathematical models resulting from different assumptions of the keratin transport, assembly and disassembly is designed to investigate the effects of the interplay between turnover and transport on the keratin organization. The collection is built as a well-designed scientific experiment by considering control and knockout of processes. To highlight and confirm the importance or existence of a given process, scenarios in which the process is absent are also considered and tested.
 Model responses are then compared to experimental observations and data published in \cite{Moch2013} to identify optimal parameter values that yield the best fit of  each of the models to the  experimental data. Finally, we identify, using an information-theoretic approach, the best scenario or model given the data and candidate models under study. 

 By employing  this reductionist phenomenological approach and systematic evaluation of  the different scenarios we not only confirm the proposed transport features of the keratin cycle and the restricted disassembly in the perinuclear region but also find that the assembly throughout the cytoplasm fit best to the experimental data. Furthermore, our particular approach allows us to demonstrate that the inward motion experimentally observed is not an emergent behavior but is an inherent property of the keratin material organization and it is due to an active transport thereby confirming recent experimental observations \cite{Leube2011,Windoffer2011}.

\section*{Methods}
\subsection*{Experimental data}
 In Moch {\it et al.} \cite{Moch2013}, the spatial distribution of the assembled keratin material in epithelial cells is measured for 15 minutes at 24 hours and 48 hours after seeding. The shape of each epithelial cell is normalized to fit a circle of fixed radius \cite{Mohl2012}. The average normalized spatial distribution is calculated over 50 cells at 24 hours (resp. 84 cells at 48 hours).   The average speed and direction of the motion of the assembled keratin material are  measured and determined at every location within the normalized cell. Finally, at every spatial location, the net assembly/disassembly is calculated. Hence, regions with preferential assembly and disassembly are identified. We will refer to regions of assembly as {\it Sources} and regions of disassembly as {\it Sinks}.   More details on the experimental data can be found in \cite{Moch2013,Herberich2011}. 

\begin{figure}[!h]
		\includegraphics[scale=1]{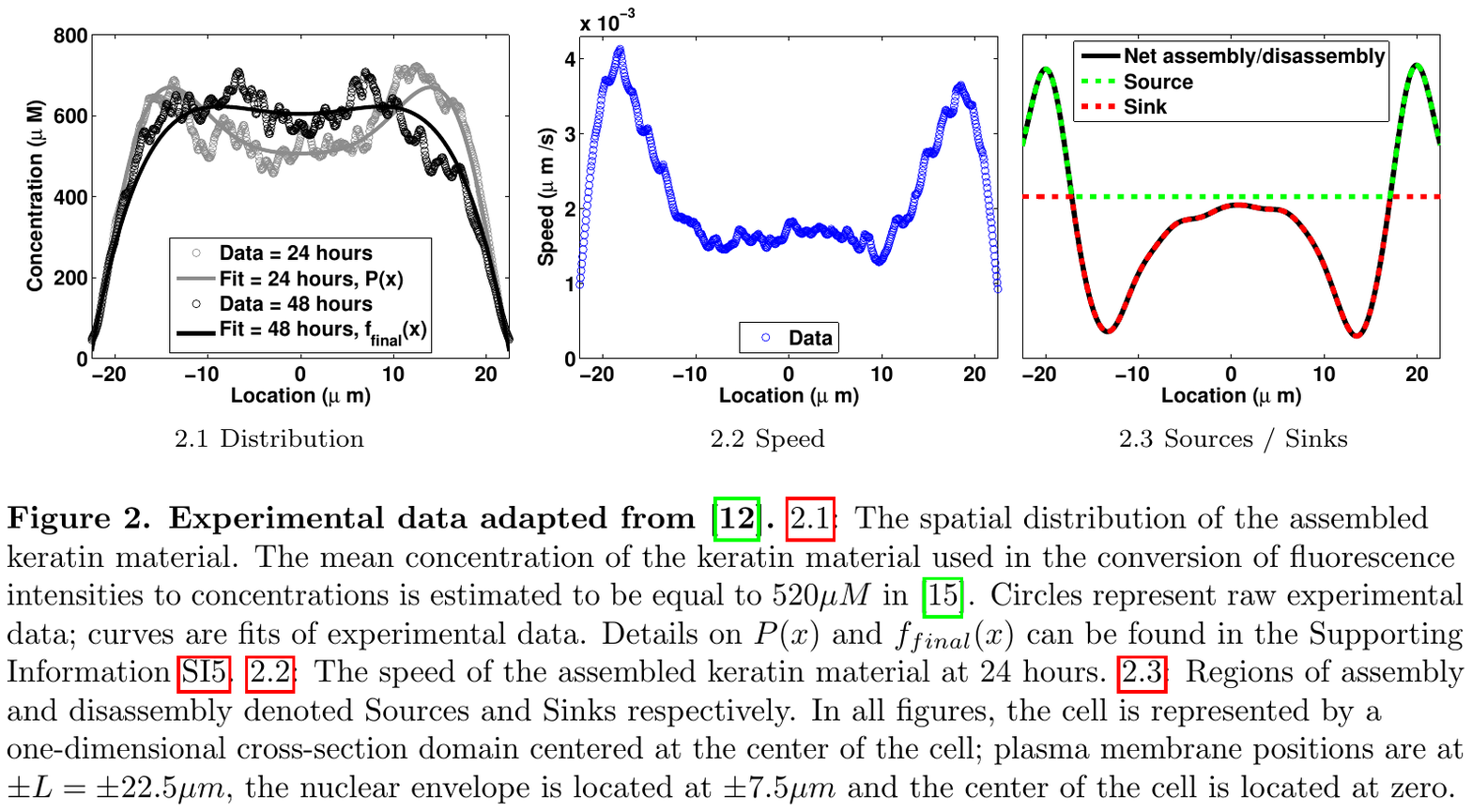}
	\caption{{\bf Experimental data adapted from \cite{Moch2013}.} 
	2.1: The spatial distribution of the assembled keratin material. The mean concentration of the keratin material used in the conversion of fluorescence intensities to concentrations is estimated to be equal to 520$\mu M$ in \cite{Feng2013}. Circles represent raw experimental data; curves are fits of experimental data. Details on $P(x)$ and $f_{final}(x)$ can be found in  \nameref{SI4}.  
	2.2:  The speed of the assembled keratin material at 24 hours. 
	2.3: Regions of assembly and disassembly denoted Sources and Sinks respectively. In all figures, the cell is represented by a one-dimensional cross-section domain centered at the center of the cell; plasma membrane positions are at $\pm L=\pm 22.5 \mu m$, the nuclear envelope is located at $\pm 7.5 \mu m$ and the center of the cell is located at zero.}\label{Figure_Data}
\end{figure}

In the present work, cells are represented as one-dimensional cross-section domains. A diameter of the normalized cell is used as the spatial domain which is centered at the center of the cell and is of length $2L$ with $L=22.5 \mu m$. Moch {\it et al.} recorded the fluorescence intensity of fluorescent protein-labeled keratins in cells. Assuming a proportionality between fluorescences and concentrations, and knowing from \cite{Feng2013} the mean concentration for keratin in keratinocytes, we convert fluorescence intensities to concentrations ($\mu M$) as follows: Concentration = Fluorescence $\times$ (Mean Concentration / Mean Fluorescence). In Fig.~\ref{Figure_Data}, the average spatial distribution, the speed of the assembled keratin, regions of assembly (Sources) and regions of disassembly (Sinks) on the one-dimensional cross-section domain are displayed.  

\subsection*{Mathematical models}
To study its organization in cells, the keratin material is categorized into  a {\bf soluble pool} composed of the soluble keratin, and an {\bf insoluble pool} representing the assembled keratin material. The state variables used to represent the soluble and insoluble pools are:
\begin{itemize}
\item $S(x,t)$ denotes the  concentration of the soluble keratin material at position $x$ at time $t$,
\item $I(x,t)$ denotes the concentration of the assembled keratin material at position $x$ at time $t$.
\end{itemize}
 From here onwards, the one-dimensional spatial domain representing the cell is defined by
$$\Omega=\{x: -L \leq x\leq L\},$$ 
where $x=0$ is the center of the cell and $x=\pm L$ are the boundary positions of the plasma membrane (Table~\ref{tab:ModelParameter}).  The general framework of the model, derived from first principles based on  experimental observations, takes into account the turnover and transport for both soluble and insoluble pools and can be stated verbally as:
\[ \begin{cases}
\textrm{Rate of change of }S=  \; \textrm{Transport of }S - \textrm{Assembly of }S\textrm{ in }I 
+ \textrm{Disassembly of }I\textrm{ in }S, \\
\textrm{Rate of change of }I=   \; \textrm{Transport of }I + \textrm{Assembly of }S\textrm{ in }I 
- \textrm{Disassembly of }I\textrm{ in }S.
\end{cases}
\]
Hence, the generalized model has the following expression; stated mathematically as:
\begin{equation}
\begin{cases}
\frac{\partial S}{\partial t}= & T_S(S) - \mathcal{A}(S) + \mathcal{D}(I),\\ \\
\frac{\partial I}{\partial t}= & T_I(I) + \mathcal{A}(S) - \mathcal{D}(I),
\end{cases}
\label{eq:GeneralFramework} 
\end{equation}
 where $T_S(\cdot)$ (resp. $T_I(\cdot)$) is the functional term describing the transport of the soluble pool (resp. insoluble pool). For example, $T_{S}(S)=-\frac{\partial }{\partial x}J_S(x,t)$ {\Big(}resp. $T_{I}(I)=-\frac{\partial }{\partial x}J_I(x,t)${\Big)} where $J_S(x,t)$ {\Big(}resp. $J_I(x,t)${\Big)} describes the flux of the soluble pool (resp. insoluble pool) at  position $x$ from the left ($x=-L$) to the right  $(x=L)$ at time $t$. The function $\mathcal{A}(S)$ (resp. $\mathcal{D}(I)$) is the assembly term (resp. disassembly term). To investigate the interplay between turnover and transport on the organization of the keratin material several assumptions are proposed for each of the functionals $T_S$, $T_I$, $\mathcal{A}$ and $\mathcal{D}$.
 \begin{table}[!h]
 \begin{adjustwidth} {-0.25in}{0in}
 	\caption{\bf{Model parameters}}
 		\begin{tabular}{|l|p{4cm}|p{4cm}|l|}
 		\hline
 			{\bf Parameter} & {\bf Definition} & {\bf Value (Unit)} & {\bf Reference} \\
 			\hline
 		$L$ &	 Half-length of spatial domain $\Omega$ & $22.5$ $(\mu m)$ & \cite{Moch2013} \\
 		&&&\\
 		$f_0(x)$ & Initial distribution of assembled keratin in cells at 24 hours & Eq.\eqref{eq:Initial} $(\mu M)$ & \cite{Moch2013} \\
 	\hline
 			$D_S$ & Diffusion coefficient of soluble pool& $0.88\pm 0.08$ $(\mu m^2/s)$ & \cite{Kolsch2010}\\
 					&&&\\
 			$D_I$ & Diffusion coefficient of insoluble pool &  $\leq 10^{-3}D_S$ $ (\mu m^2/s)$  & \\
 					&&&\\
 			$u$ & Speed of insoluble pool &  $0.002 - 0.008$ $ (\mu m / s) $  & \cite{Moch2013,Woll2005,Kolsch2009,Windoffer1999}\\
 						&&&\\
 						$v(x)$ & Almost constant speed of active transport of insoluble pool &  Eq.\eqref{eq:ConstantSpeed} ($ \mu m / s$) & \cite{Windoffer1999,Moch2013}\\
 				&&&\\
 			$u(x)$ & Variable speed of active transport of insoluble pool &  Eq.\eqref{eq:estimatev(x)WholeCellSymmetrical} ($ \mu m / s$) & \cite{Moch2013}\\
 		\hline
 			$k_{ass}$ & Rate of assembly of soluble pool &  TBD (linear model: $s^{-1}$, nonlinear model: $\mu M.s^{-1}$) &  \\
 			&&&\\
 			$k_{ass}(x)$ & Localized rate of assembly of soluble pool &   Eq.\eqref{eq:SourceKAss} with $k_{ass}$ TBD (linear model: $s^{-1}$, nonlinear model: $\mu M.s^{-1}$) &  \\
 						&&&\\ 				
 		    					$k_S$ & Concentration for half-saturation of assembly rate (nonlinear model) & TBD ($\mu M$) &  \\
 &&&\\
 			$k_{dis}$ & Rate of disassembly of insoluble pool into soluble pool & TBD (linear model: $s^{-1}$, nonlinear model: $\mu M.s^{-1}$) &  \\
 			&&&\\
 		    $k_{dis}(x)$ & Localized rate of disassembly of insoluble into soluble pool & Eq.\eqref{eq:SinkKDis} or \eqref{eq:LocDisWhole} with $k_{dis}$ TBD (linear model: $s^{-1}$, nonlinear model: $\mu M.s^{-1}$)  &  \\
 						&&&\\
 					$k_I$ & Concentration for half-saturation for disassembly rate (nonlinear model) & TBD ($\mu M$) &  \\\hline
 		\end{tabular}
 	\begin{flushleft}
 	The model parameters used in all the scenarios. (TBD = To Be Determined by fitting model solutions to experimental data).
 	\end{flushleft}
 	\label{tab:ModelParameter}
 	\end{adjustwidth}
 \end{table}
 \newpage

\begin{figure}[!h]
		\includegraphics[scale=.5]{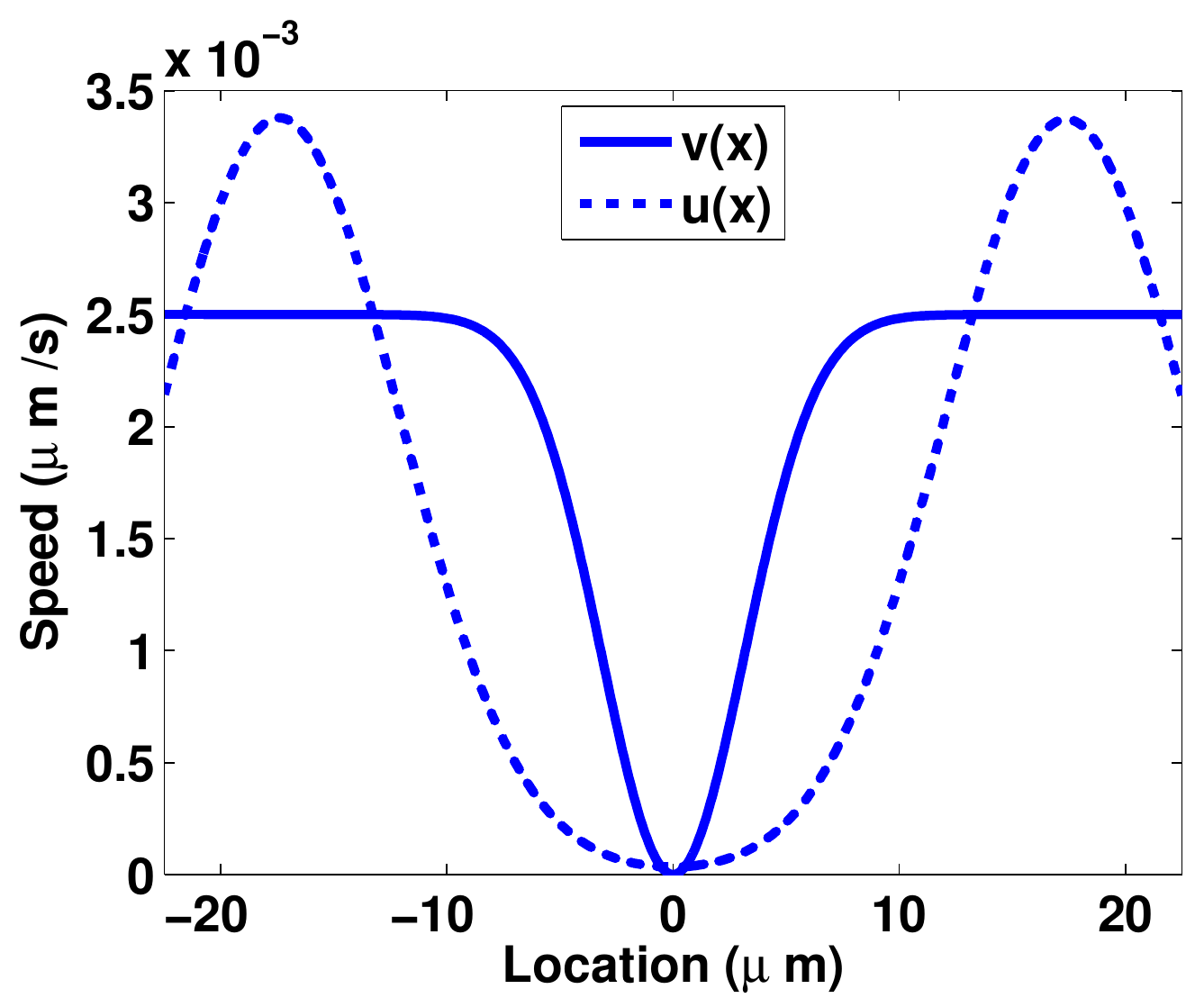}
\caption{{\bf Profiles of the active transport speeds $u(x)$ and $v(x)$ for the insoluble pool} both of which are derived from the experimental measurements in \cite{Moch2013}.  The function $u(x)$ is derived from the profile of speeds and $v(x)$ is approximated as the average value of these speeds. Details on the derivation of $u(x)$ and $v(x)$ are given in \nameref{SI1}.}
\label{Figure_AllSpeed}
\end{figure}

\paragraph{Modes of transport}
Molecules of the soluble pool are assumed to be subjected to the Brownian motion; the soluble pool is diffusible with a diffusion coefficient $D_S$. The functional term of transport for the soluble pool is given by
\begin{align}
T_S(S)&:=D_S\frac{\partial ^2 S}{\partial x^2}.
\label{eq:TransportS}
\end{align}
Passive transport for both pools is assumed in all models. Diffusion is assumed for the insoluble pool to describe the wiggling motion of the keratin filaments in cells.  The diffusion coefficient of the insoluble pool is set to be much smaller than that of the soluble pool: $0<D_I\leq 10^{-3}D_S$. It is assumed that only the insoluble pool can be driven by an active transport. Experimental evidence show that the assembled intermediate filament proteins in the form of filament precursors, squiggles or filaments move along microtubules and actin filaments by interacting via molecular motors \cite{Helfand2004,Robert2014,Yoon2001,Windoffer1999,Windoffer2011}. An  active transport (also called the inward drift) of the insoluble pool from the plasma membrane to the center of the cell is hypothesized based on reports of the motion of the assembled keratin material mostly towards the nucleus in epithelial cells \cite{Windoffer1999,Windoffer2011,Kolsch2009}. The speed of the active transport  is set to be almost constant  $v(x)$ everywhere or variable $u(x)$ everywhere (Fig.~\ref{Figure_AllSpeed}).  Based on experimental observations, both speeds are assumed to decay around the nucleus towards the center of the cell. The variable speed $u(x)$ is estimated using the profile of  the average speeds measured in \cite{Moch2013} (Fig.~\ref{Figure_Data}.2). The magnitude of the almost constant speed $v(x)$ is set to be the average value of the variable speed over the cell. Details on the derivation of the estimates of $v(x)$ and $u(x)$ are given in  \nameref{SI1}. Hence, the functional term for the transport of the insoluble pool can take three different forms:
\begin{align}
T_I(I)&:=\left \{
\begin{array}{ll}
D_I\frac{\partial ^2 I}{\partial x^2},& \textrm{no drift,}\\ \\
D_I\frac{\partial ^2 I}{\partial x^2}+ sgn(x)v(x)\frac{\partial I}{\partial x}, & \textrm{inward drift with almost constant speed,}\\ \\
D_I\frac{\partial ^2 I}{\partial x^2}+ sgn(x)u(x)\frac{\partial I}{\partial x}, & \textrm{inward drift with variable speed,}
\end{array}
\label{eq:TransportI}
\right .
\end{align}
where $v(x)$ is given in \eqref{eq:ConstantSpeed} and $u(x)$ in \eqref{eq:estimatev(x)WholeCellSymmetrical};   the two functions representing the speeds of the active transport are graphed in Fig.~\ref{Figure_AllSpeed}. The function $sgn(x)$  defined by
$$sgn(x)=\left \{\begin{array}{rr}
1 & x>0,\\ \\
-1 & x<0,
\end{array}\right.$$
describes the inward direction of the active transport at any location of the spatial domain $\Omega$ centered at zero.

Combining the modes of transport for the soluble and insoluble pools,  three types of transport are considered for the keratin material in the epithelial cell.

\paragraph{Expressions of assembly / disassembly reactions}
In the present work, the turnover is composed of two reactions: the assembly / aggregation / polymerization of units of the soluble pool to grow the insoluble pool and the disassembly / solubilization / depolymerization of the insoluble pool into units of soluble pool. It is assumed that the assembly process is a function of the soluble pool only, whereas the disassembly process is a function of the insoluble pool only. The simplest case is to consider a linear model that assumes linear exchanges between the two pools.  The linear model can be stated as follows:
\begin{align}
\pm \Big ( \mathcal{A}(S) - \mathcal{D}(I) \Big )& := \pm \Big (k_{ass}(\cdot)S -k_{dis}(\cdot)I\Big),
\label{eq:LinearModel}
\end{align} 
where $k_{ass}(\cdot)$ (resp. $k_{dis}(\cdot)$) is the rate of assembly of the soluble pool  (resp. disassembly of the insoluble pool). Both rates can either be constant, $k_{ass}$ and $k_{dis}$, or space-dependent $k_{ass}(x)$ and $k_{dis}(x)$. 

In the second case, the turnover is assumed to depend on the enzymatic activities \cite{Snider2014}. For instance, the solubilization of the insoluble pool into soluble proteins (disassembly) is triggered by a kinase activity and the assembly of insoluble pool is induced by the dephosphorylation of soluble proteins by a phosphatase \cite{Inagaki2006}. The turnover term is assumed to be of the Michaelis-Menten form stated as: 
\begin{align}
\pm \Big ( \mathcal{A}(S) - \mathcal{D}(I) \Big )& := \pm \left (\frac{k_{ass}(\cdot)S}{k_S+S} -  \frac{k_{dis}(\cdot)I}{k_I+I}\right),
\label{eq:NonLinearModel}
\end{align} 
where $k_{ass}(\cdot)$ (resp. $k_{dis}(\cdot)$) is the maximum rate of assembly of the soluble pool  (resp. disassembly of insoluble pool) and $k_S$ (resp. $k_I$) is the concentration at which the assembly (resp. disassembly) rate is half of $k_{ass}(\cdot)$ (resp. $k_{dis}(\cdot)$). When Michealis-Menten dynamics are used, the model is called  nonlinear. Both rates can either be constant or space-dependent describing the intracellular localization of the post-translational modification enzymes. 

\paragraph{Profiles of assembly and disassembly rates}
 As previously mentioned, the assembly and disassembly rates, $k_{ass}(\cdot)$ and $k_{dis}(\cdot)$, used in the linear and nonlinear models can be constant or space-dependent functions.  The profile (shape) of the space-dependent function $k_{ass}(x)$ is derived from the spatial profile of regions of assembly (Sources) measured in \cite{Moch2013} (Fig.~\ref{Figure_Data}.3). Details of the derivation of the Sources $k_{ass}(x)$ are given in \nameref{SI2}. 

 Two types of shapes for $k_{dis}(x)$ are assumed to represent two types of  localization of the disassembly in cells. First, similarly to the assembly rate, the profile of the disassembly rate is deduced from the experimental data published in \cite{Moch2013} (Fig.~\ref{Figure_Data}.3); the spatial profile of the disassembly regions,  Sinks, is used to build the shape of the first space-dependent disassembly rate. This disassembly rate is called of type Sinks. Second, it is assumed that disassembly is localized around the nucleus; a mollified step-function is designed to describe this assumption. This second space-dependent disassembly rate is called of type Mollify. Details on the derivation of the two $k_{dis}(x)$ rates are given in \nameref{SI3}. 

\begin{figure}[!h]
			\includegraphics[scale=1]{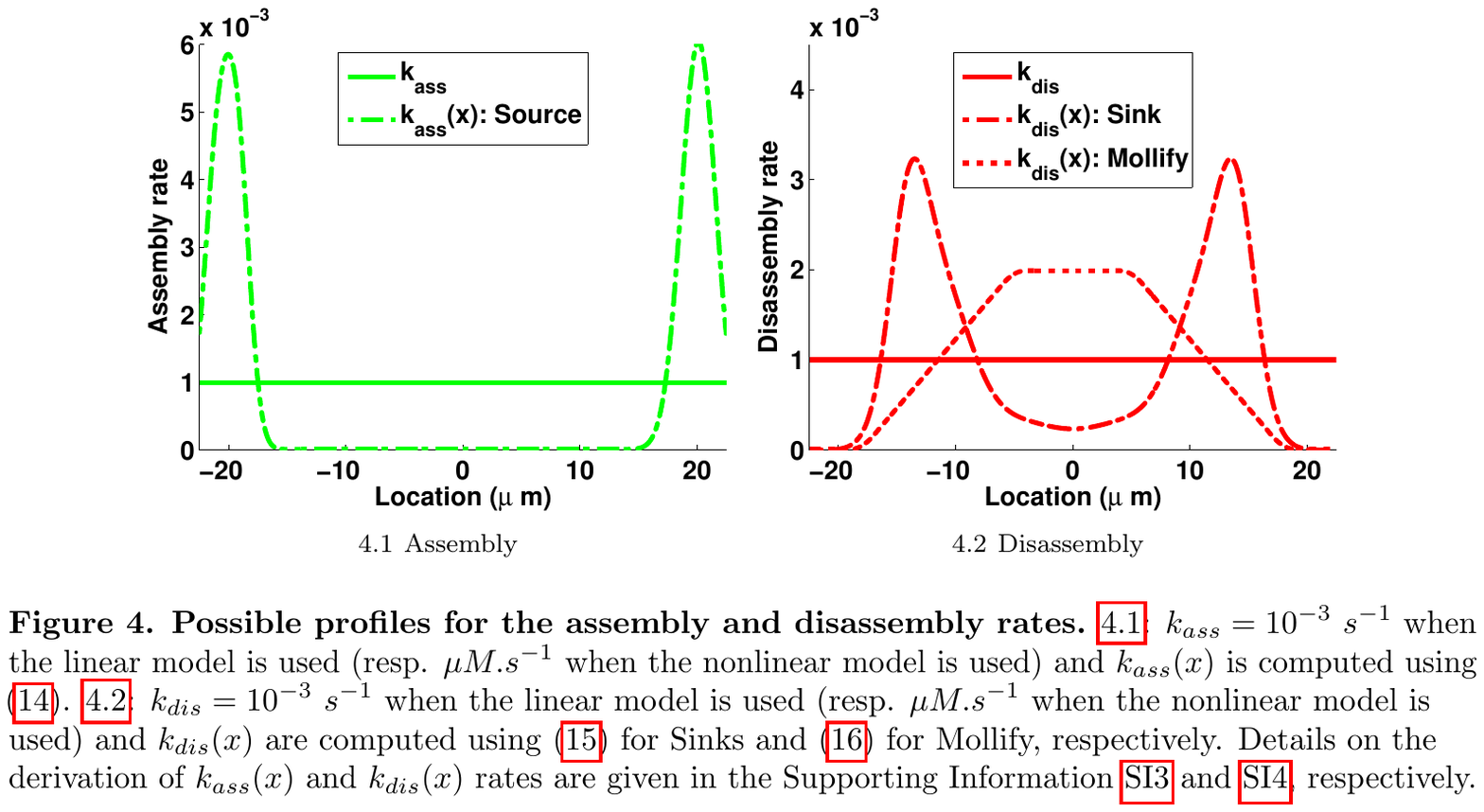}
\caption{{\bf Possible profiles for the assembly and disassembly rates.} 
4.1: $k_{ass}=10^{-3}$ $s^{-1}$ when the linear model is used (resp. $\mu M.s^{-1}$ when the nonlinear model is used) and $k_{ass}(x)$ is computed using \eqref{eq:SourceKAss}. 
4.2: $k_{dis}=10^{-3}$  $s^{-1}$ when the linear model is used (resp. $\mu M.s^{-1}$ when the nonlinear model is used) and $k_{dis}(x)$ are computed using \eqref{eq:SinkKDis} for Sinks and \eqref{eq:LocDisWhole} for Mollify, respectively. Details on the derivation of $k_{ass}(x)$ and $k_{dis}(x)$ rates are given in \nameref{SI2} and \nameref{SI3}, respectively. Parameter values in \eqref{eq:SourceKAss}, \eqref{eq:SinkKDis} and \eqref{eq:LocDisWhole} are chosen in such a way that their profiles give the same total amount of assembly / disassembly over the spatial domain $\Omega$. }
\label{fig:RateProfile}
\end{figure}

For the sake of illustration, the two profiles of $k_{ass}(\cdot)$ and the three profiles of $k_{dis}(\cdot)$ are shown in Fig.~\ref{fig:RateProfile}. 

\begin{figure}[!h]
	\includegraphics[scale=1]{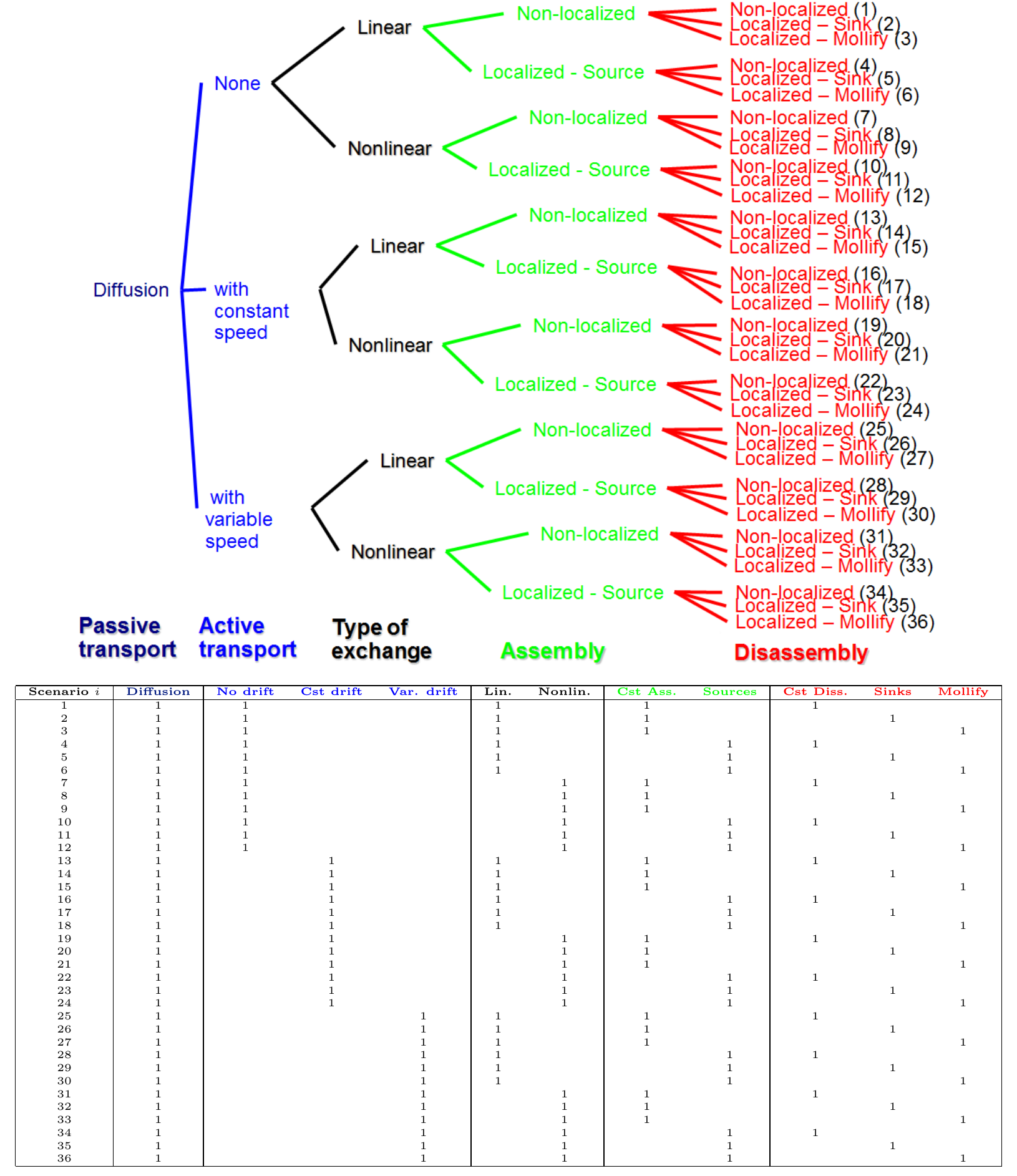}
	\caption{{\bf The 36 scenarios to be considered.} Top: Number in parentheses is the scenario index $i$. Bottom: The numerical value 1 denotes that the process of interest is in Scenario $i$.}
	\label{fig:Hypotheses}
\end{figure}
\begin{figure}[!h]
		\includegraphics[scale=0.5]{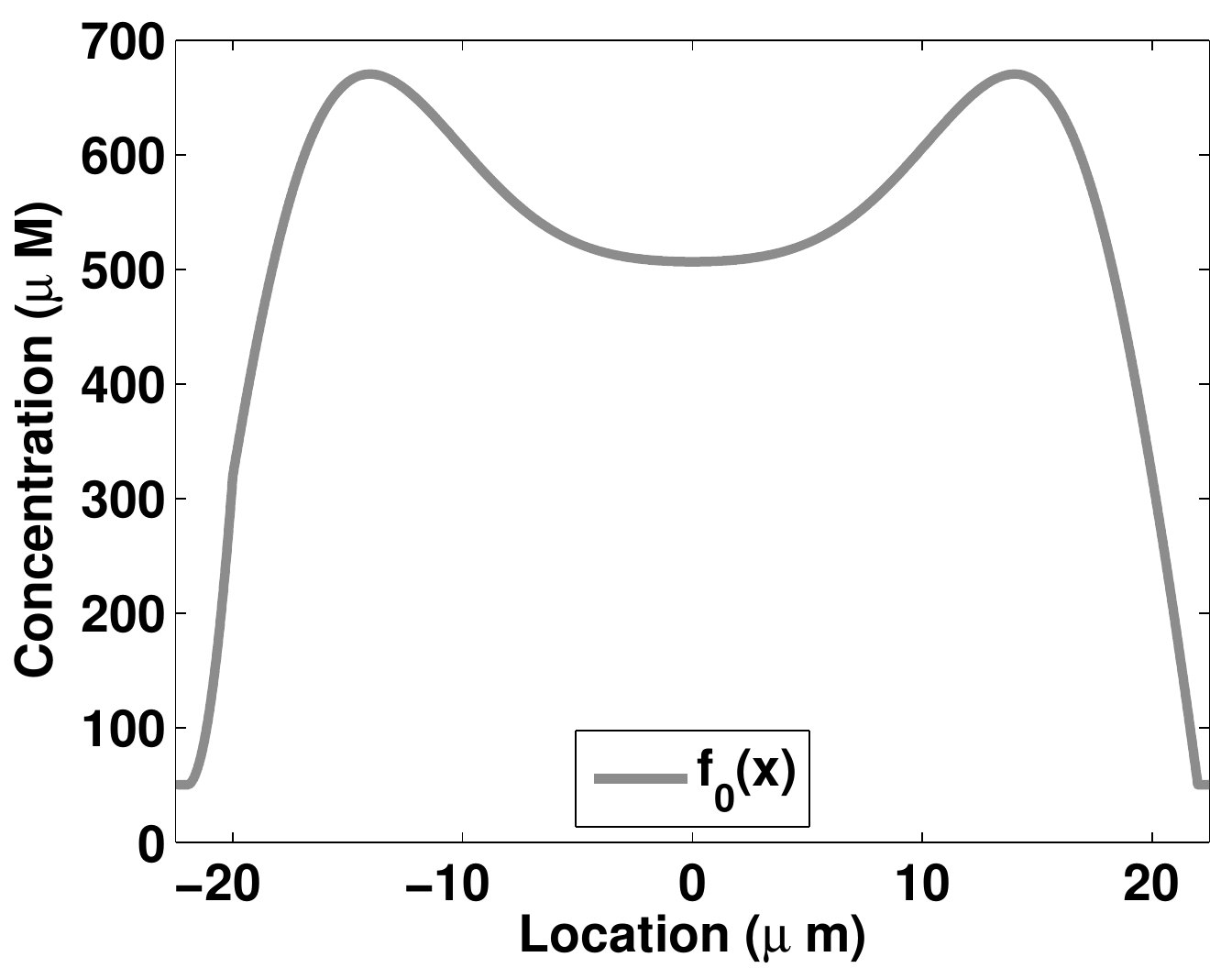}
	\caption{{\bf Initial profile of the insoluble pool} $f_0(x)$ defined in \eqref{eq:Initial}.}
	\label{fig:Initial_Portet}
\end{figure}

Accounting for the three modes of transport and considering the turnover described by either linear and nonlinear models with 6 possible combinations of the profiles for the assembly and disassembly rates, a collection of 36 scenarios (models) is defined; each scenario follows the general form stated by system \eqref{eq:GeneralFramework}. Details of the 36 scenarios are given in Fig.~\ref{fig:Hypotheses}. All scenarios are considered with the same  initial conditions given by
\begin{equation}
\begin{cases}
S(x,t_0)&=\frac{0.05}{0.95}f_0(x), \\ \\
I(x,t_0)&=f_0(x),
\end{cases}
 \qquad \text{for all $x \in \Omega$},
\label{eq:IC}
\end{equation}
where $f_0(x)$, defined in \eqref{eq:Initial}, is chosen to be a mollified version of the profile of the assembled keratin at time $t_0=24$ hours averaged over all the normalized cells (Fig.~\ref{fig:Initial_Portet}).  Details of the derivation of $f_0(x)$ are given in \nameref{SI4}. Initial conditions describe the observed fact that the soluble pool (resp. insoluble pool) represents $5\%$ (resp. $95\%$) of the total keratin material \cite{Chou1993}. All scenarios are considered with boundary conditions describing the impermeability of the plasma membrane for the keratin material
 \begin{align}
 J_S(\pm L, t)= 0 =  J_I(\pm L, t), \qquad t\geq t_0,
 \label{eq:BC}
 \end{align}
where $J_S$ is the flux of the soluble pool and $J_I$ is the flux of the insoluble pool as defined below system \eqref{eq:GeneralFramework}. The model parameters used in all the scenarios are listed in Table~\ref{tab:ModelParameter}.

\subsection*{Comparison between mathematical models and experimental data}
\paragraph{Parameter estimation}
Let $p$ denote the set of all the model parameters for each scenario. To estimate the  optimal set of parameter values $p$ for each scenario, the solution of each scenario is compared to experimental data using the following objective function (an estimation of the distance between the experimental data and the model response):
\begin{align}
\Psi(p)=\sum_{i=1}^{N_x} \Big[ I(x_i,t_{final},p)-f_{final}(x_i) \Big]^2,
\end{align}
where $t_{final}$ is equal to $48$ hours. The experimental data at $t_{final}$ is represented by $f_{final}(x)$ and the model solution at $t_{final}$ of the considered scenario evaluated with the parameter values $p$ is $I(x,t_{final},p)$. To obtain the  model solution for each scenario, the corresponding system is numerically integrated from $t_0=24$ hours to $50$ hours using the MATLAB solver for partial differential equations, pdepe \cite{MATLAB:2012}. Solutions are computed at $N_x$ locations of the spatial domain $\Omega$ with $N_x=200$. To  carry out computations, the raw data in which the concentration of the assembled keratin material is known at 623 spatial locations is approximated by $f_{final}(x)$ defined in \nameref{SI4} by \eqref{eq:Final}; $f_{final}(x)$ is the fit of the average profile of the assembled keratin measured after $48$ hours on the normalized cells (see the black profile in Fig.~\ref{Figure_Data}.1).

The estimation of the parameter values for the 36 scenarios, i.e. the determination of parameter values $\hat p$ that provide the best fit to the  experimental data, is done by minimizing the objective function, $\Psi(p)$, such that 
\begin{align}
\Psi(\hat p)=\min_{p}\Psi(p).
\label{eq:Min}
\end{align}
Minimization of the objective function is done by a parallelizable genetic algorithm described in \cite{Dorsey1995}.

In this study, only constant parameter values of the turnover reactions are optimized. When the linear model is considered, for each scenario, two parameters $k_{ass}$ and $k_{dis}$ are estimated. When the nonlinear model is considered, four parameters, $k_{ass}$, $k_{S}$, $k_{dis}$ and $k_{I}$,  are estimated for each scenario. In all scenarios, the diffusion coefficients for the soluble and insoluble pools as well as the active transport speeds $v(x)$ and $u(x)$  are considered as fixed/known parameters or functions and are set to values measured in previous studies  \cite{Moch2013,Kolsch2010,Windoffer2011} (Table~\ref{tab:ModelParameter}). The diffusion coefficients are taken as $D_S=0.88\mu m ^2 /s$ and $D_I=9.5\times 10^{-4}D_S$, and the ``constant speed'' $u$ in $v(x)$ as defined in \eqref{eq:ConstantSpeed} is set to be $u=0.0025 \mu m/ s$ \cite{Kolsch2010,Windoffer1999}.

\paragraph{Model selection}
In order to select the best model out of the 36 scenarios, we use an information-theoretic approach. Specifically, we employ the Akaike information criterion  (AIC) \cite{Johnson2004,Burnham2002} to select from all the scenarios, the best model that best captures experimental observations given the collection of models considered in this study. Since we use the Least Squares principle to estimate the values of the constant free parameters (2 free parameters for linear models and 4 for nonlinear models), the Akaike criterion for Scenario $i$, $AIC_i$, takes the following form:
\begin{align}
AIC_i= N_x \ln \left ( \hat{\sigma}^2\right)+2K,
\label{eq:Akaike}
\end{align}
 where $\hat{\sigma}^2=\frac{\Psi _i(\hat{p})}{N_x}$ is the estimate of the variance with $\Psi _i(\hat{p})$ the residual \eqref{eq:Min} estimated for Scenario $i$ and $N_x$ the size of the sample ($N_x$=200). $K$ is the number of estimated parameters; {\it i.e.} the number of free parameters in Scenario $i$ plus one for the estimate of the variance ($K=3$ for linear model and $K=5$ for nonlinear models). The scenario with the lowest AIC value is the best model. The AIC selects a model with the least number of parameters that best fits experimental observations. To rank and compare scenarios the Akaike weights $w_i$  are calculated and  these are known as the weights of evidence in favor of Scenario $i$ being the actual best model  given the experimental data and the collection of scenarios considered. The Akaike weights are expressed as follows:
 \begin{align}
 w_i=\frac{exp(-1/2\Delta_i)}{\sum_{r=1}^R exp(-1/2\Delta_r)}, \quad \textrm{with} \quad \Delta_i=AIC_i-\min_i AIC_{i}
 \label{eq:AkaikeWeight}
 \end{align}
 where $R$ represents the number of scenarios considered ($R=36$) and $\Delta_i$ being the difference in AIC with respect to the AIC of the preferred scenario $\min_i AIC_{i}=AIC_{min}$. It must be noted that the Akaike weights $w_i$ sum to 1 and are interpreted as the probability that Scenario $i$ is the best model  given the experimental data and the collection of scenarios considered. The models, ranked from the largest to the smallest Akaike weights, whose Akaike weights sum to 0.95 form the confidence set of the models that captures, more faithfully,  experimental data. The ratio of the Akaike weights $w_i/w_j$ (also known as the evidence ratio) is used to compare model pairs. Furthermore, the relative importance of a process can be estimated by summing the Akaike weights of all scenarios involving the process of interest. We will denote by $w_*^+$ the sum of the Akaike weights of the scenarios including the process of type $*$. This sum can be interpreted as the probability that the process of type $*$ is the best type of the process given the experimental data and the collection of models considered.

\section*{Results}
\subsection*{Best scenario}
We employ a two-step process in order to find the best scenario. First, the best fit for each scenario is found by minimizing the objective function \eqref{eq:Min}. Secondly, the Akaike information criterion \eqref{eq:Akaike} is used to select the best of the best fits out of all the scenarios.

\begin{sidewaysfigure}[h]
	\includegraphics[scale=1]{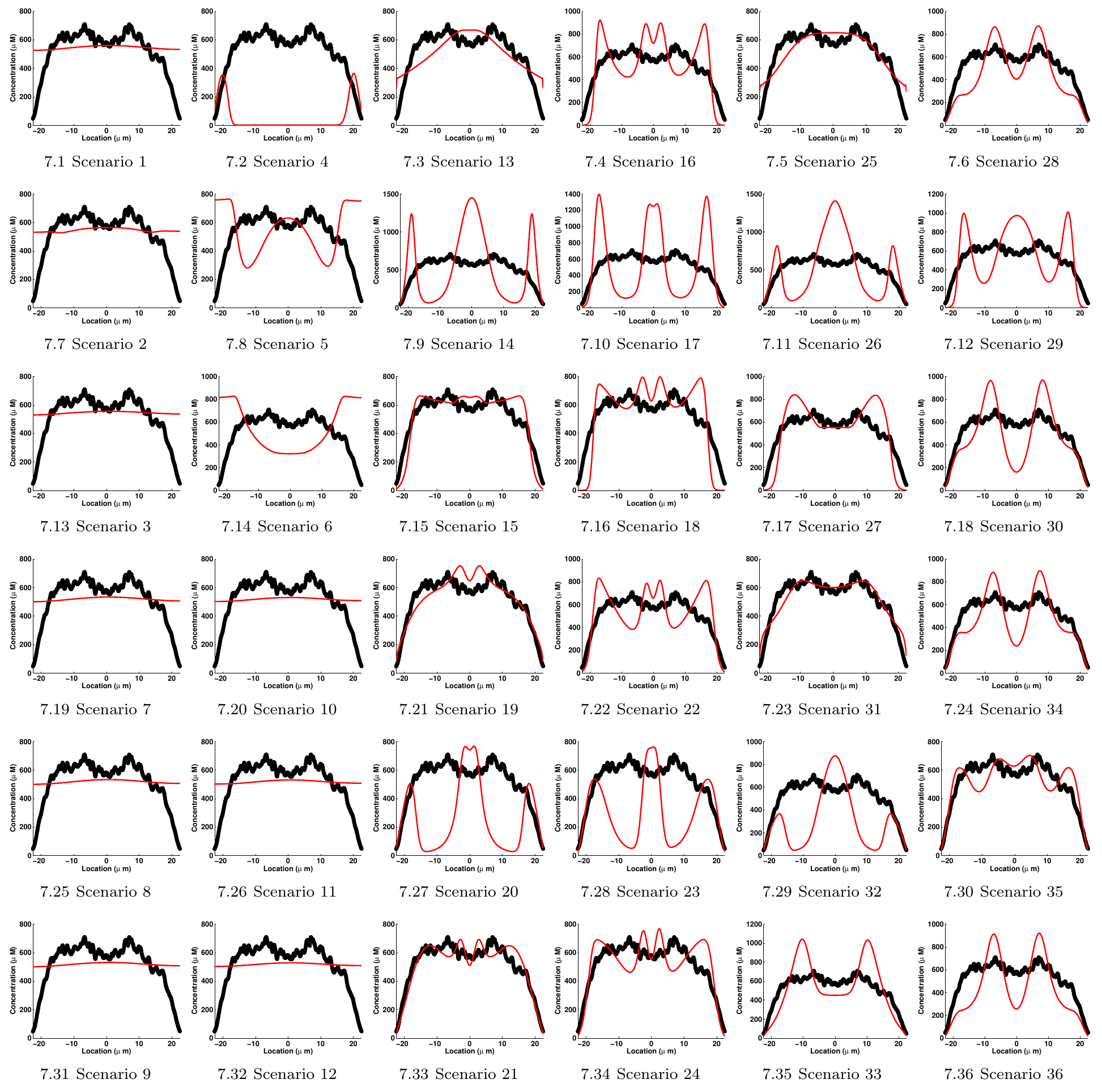}
	\caption{{\bf Best fit for each of the 36 scenarios.} Red curve is the response $I_i(x,t_{final},\hat{p})$ at $48$ hours of Scenario $i$. Black circles are data at $48$ hours as in Fig.~\ref{Figure_Data}.1.}
	\label{fig:NonL}
\end{sidewaysfigure}

The best fit for each of the 36 scenarios is presented in Fig.~\ref{fig:NonL} in the following order:
\begin{enumerate}
\item Row-wise: 
\begin{itemize}
\item Scenarios in the first three rows (1 to 3) have linear terms for turnover.
\item Scenarios in the  last three rows (4 to 6) have Michaelis-Menten type turnover terms. 
\item Rows 1 and 4 are scenarios with constant disassembly rate.
\item Rows 2 and 5 are scenarios with localized disassembly rate of type Sinks. 
\item Rows 3 and 6 have scenarios with localized disassembly rate of type Mollify. 
\end{itemize}
\item Column-wise: 
\begin{itemize}
\item The first two columns (1 and 2) include scenarios with diffusion alone without drift. 
\item Columns 3 and 4 display scenarios with drift with an almost constant speed. 
\item The last 2 columns (5 and 6) are scenarios with drift with variable speed. 
\item Odd columns (1, 3 and 5) are scenarios with constant assembly rate. 
\item Even columns (2, 4 and 6) are scenarios with localized assembly rate of type Sources.
\end{itemize}
\end{enumerate}

According to the $AIC_i$ values, the best model, the best of the best fits, is identified as being Scenario 21, $AIC_{21}=\min_{i}AIC_{i}$ ($3^{rd}$ column of Table~\ref{Table:Akaike}). Since Scenario 21's Akaike weight is over $0.95$ ($5^{th}$ column of Table~\ref{Table:Akaike}), it is the only model out of the set considered that satisfies the confidence criterion; Scenario 21 matches very well experimental observations and  data. The second best model is Scenario 31. The evidence ratio of scenarios 21 and 31 $w_{21}/w_{31}$ is equal to $36.67\times 10^6$; {\it i.e.} Scenario 21 is about 36 millions times more likely than Scenario 31 to be the best model given the experimental data and the collection of models considered. We consider this to be strong evidence in support of Scenario 21. 
\begin{table}[!h]
\begin{adjustwidth}{-0.25in}{0in}
	\caption{\bf{Results of the model selection for the best fit of each of the 36 scenarios.}}
	\begin{tabular}{|l|l|l|l|l|l|}
		\hline
		\textbf{Scenario $i$}&\textbf{$K$}&\textbf{$AIC_i$}&\textbf{$\Delta_i$}&\textbf{$w_i$}&{\bf Rank}\\\hline
		1&3&2012.996&510.3519&0&13\\\hline
		2&3&2019.144&516.4999&0&19\\\hline
		3&3&2018.634&515.9895&0&18\\\hline
		4&3&2512.249&1009.604&0&36\\\hline
		5&3&2283.252&780.6081&0&28\\\hline
		6&3&2340.069&837.4254&0&31\\\hline
		7&5&2011.436&508.7923&0&11\\\hline
		8&5&2011.724&509.0804&0&12\\\hline
		9&5&2013.609&510.9647&0&14\\\hline
		10&5&2014.842&512.1975&0&15\\\hline
		11&5&2015.148&512.5043&0&16\\\hline
		12&5&2017.034&514.3897&0&17\\\hline
		13&3&1839.930&337.2861&0&7\\\hline
		14&3&2479.026&976.3817&0&35\\\hline
		15&3&1738.738&236.0939&0&5\\\hline
		16&3&2114.919&612.2750&0&23\\\hline
		17&3&2458.497&955.8532&0&34\\\hline
		18&3&2002.972&500.3279&0&9\\\hline
		19&5&1668.771&166.1274&$0$&3\\\hline
		20&5&2387.721&885.0770&0&32\\\hline
		21&5&1502.644&0&0.99999&1\\\hline
		22&5&2073.809&571.1653&0&21\\\hline
		23&5&2338.356&835.7119&0&29\\\hline
		24&5&1893.875&391.2309&0&8\\\hline
		25&3&1703.773&201.1292&$0$&4\\\hline
		26&3&2410.565&907.9213&0&33\\\hline
		27&3&2006.784&504.1403&0&10\\\hline
		28&3&2067.488&564.8442&0&20\\\hline
		29&3&2243.066&740.4220&0&27\\\hline
		30&3&2145.639&642.9953&0&26\\\hline
		31&5&1537.479&34.83477&$2.7\times10^{-8}$&2\\\hline
		32&5&2339.200&836.5561&0&30\\\hline
		33&5&2119.447&616.8033&0&24\\\hline
		34&5&2074.025&571.3811&0&22\\\hline
		35&5&1834.186&331.5420&0&6\\\hline
		36&5&2143.595&640.9505&0&25\\\hline
		\end{tabular}
	\begin{flushleft}
	{The numerical value $0$ in the $5^{th}$ column denotes a value lower than $10^{-12}$. In the $6^{th}$ column, ``Rank'' denotes the ranking of scenarios, {\it i.e. } the descending order of Akaike weights $w_i$. The use of weights allows the "quantitative" comparison of the adequacy of scenarios because of the normalization to 1.}
	\end{flushleft}
	\label{Table:Akaike}
	\end{adjustwidth}
\end{table}
\newpage

\begin{figure}[!h]
	\includegraphics[scale=1]{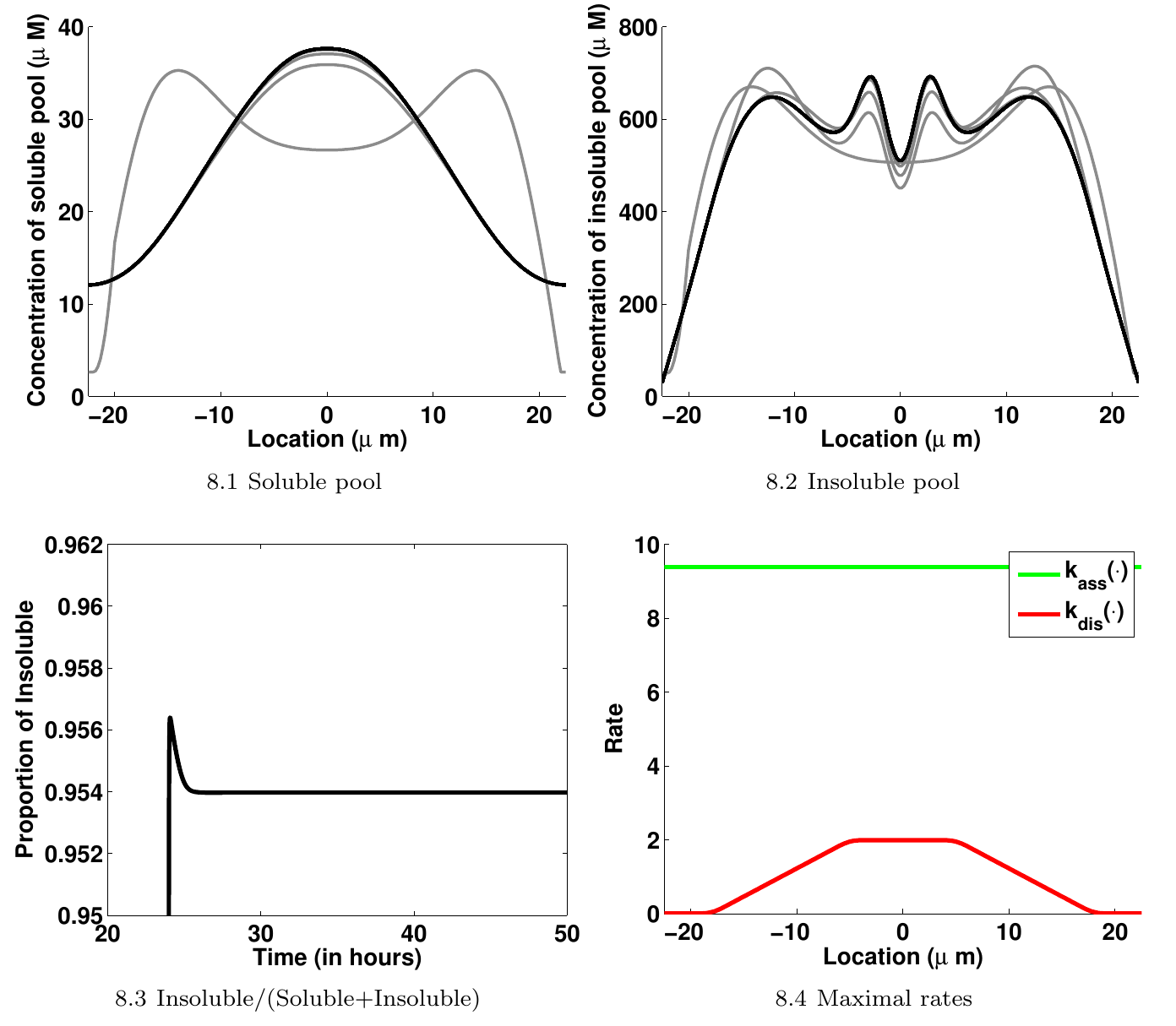}
	\caption{{\bf Details about the best model Scenario 21.} 
	8.1: Snapshots from 24 hours (gray) to 48 hours (black) taken every 30 minutes of the profile of the soluble pool over the cell. 
	8.2: Snapshots from 24 hours (gray) to 48 hours (black) taken every 30 minutes of the profile of the insoluble pool over the cell. 
	8.3:  The distribution of the keratin material between the soluble and insoluble pools over time. 
	8.4: The profiles of assembly and disassembly maximal rates used with  the Michaelis-Menten type turnover terms in Scenario 21. The estimates are  $k_{ass}=9.3819\mu M/s$, $k_S=570.73\mu M$, $k_{dis}=0.9998\mu M/s$ leading to $k_{max}=1.976 \mu M/s$ in \eqref{eq:LocDisWhole} and $k_I=976.07\mu M/s$. }
	\label{fig:Scenario21}
\end{figure}

Scenario 21 is characterized by an inward drift with an almost constant speed for the insoluble pool, turnover terms of Michaelis-Menten type, a constant assembly rate and a disassembly rate of type Mollify.  The profiles of assembly and disassembly rates are displayed in Fig.~\ref{fig:Scenario21}. For this scenario, the estimated optimal parameter values are $k_{ass}=9.3819\mu M/s$, $k_S=570.73\mu M$, $k_{dis}=0.9998\mu M/s$ leading to $k_{max}=1.976 \mu M/s$ in \eqref{eq:LocDisWhole} and $k_I=976.07\mu M$. Based on the Michaelis-Menten constants for the assembly and disassembly processes, since $k_S<k_I$, an enzyme that would be involved in the solubilization of the assembled keratin material requires a higher substrate concentration to achieve a given reaction speed than an enzyme that would be involved in the assembly of soluble proteins. 
Snapshots of the soluble and insoluble pool profiles taken every 30 minutes from $t_0$ to $t_{final}$ are displayed in Fig.~\ref{fig:Scenario21}. After only 2 hours, the solution stabilizes to its final profile. Scenario 21 preserves the repartition of the keratin material between the soluble and insoluble pools over time, about $95\%$ of the keratin material is assembled to form the insoluble pool. Finally, the characteristic time scales of the passive transport ($\tau_{Diffusion}$),  active transport ($\tau_{Drift}$) and turnover ($\tau_{Reaction}$) for Scenario 21 are estimated by using an adimensionalization of system \eqref{eq:GeneralFramework} corresponding to Scenario 21. Details on calculations are given in \nameref{SI5}. The time scales of the processes included in Scenario 21 are ordered as follows:
\[ \tau_{Reaction}(10^2s)<\tau_{Diffusion}^S (10^3s)<\tau_{Drift}(10^4s)<\tau_{Diffusion}^I (10^6s). \]
Using Peclet's number ($Pe = \tau_{Diffusion}/\tau_{Drift}\gg 1$), it is found that the transport of  the assembled keratin material by drift is faster than by diffusion, the dominant mode of transport is the active transport. Using Damk\"ohler number ($Da=\tau_{Drift}/\tau_{Reaction}\gg 1$), it is found that the  active transport time scale is greater than the reactive time scale; overall, Scenario 21 is controlled by active transport. 

The complete ranking based on the Akaike weights $\omega_i$ of all the scenarios is given in the $6^{th}$ column of Table~\ref{Table:Akaike}. 
It is worthwhile remarking that differences in $AIC_i$, $\Delta_i$, could have been used for ranking purposes; in this case, we would have obtained the same ranking. The rationale of using $\omega_i$ is that it allows us to quantify how preferably each candidate model is via the normalization to 1.

\subsection*{Importance of the type of process}
 Using the sums of the Akaike weights, $w_*^+$, we investigate several questions to evaluate the relative importance of the different types considered for a process (Tables \ref{Table:Importance} and \ref{Table:AssemblyDisassembly}). The set of the 36 scenarios is partitioned into categories with respect to the types of  processes considered. For instance, considering the transport process, the collection of the 36 scenarios is partitioned into 3 categories (Fig.~\ref{fig:Hypotheses}): scenarios with no drift (Scenarios from 1 to 12), scenarios with an almost constant speed drift (Scenarios from 13 to 24) and  scenarios with a variable speed drift (Scenarios from 25 to 36). The sum of the Akaike weights of each category is then  calculated, compared and ordered to characterize which type is more likely to be present. In what follows the sign $>$ denotes the relative importance, measured in probabilistic terms, of the type of process.
 \begin{table}[!h]
 \begin{adjustwidth}{-1.25in}{0in} 
 \caption{
 {\bf Importance of the type of process.}}
 \begin{tabular}{|l|l|l||l|l||l|l||l|l|l|}
 \hline
 \multicolumn{3}{|l||}{\bf Type of Transport} & \multicolumn{2}{|l||}{\bf Type of Turnover Term}& \multicolumn{2}{|l||}{\bf Type of Assembly}& \multicolumn{3}{|l|}{\bf Type of Disassembly}\\ \hline
 \textbf{$w_{NoDrift}^+$}&\textbf{$w_{CstDrift}^+$}&\textbf{$w_{VarDrift}^+$}& 	\textbf{$w_{Linear}^+$}&\textbf{$w_{Nonlinear}^+$}& 	\textbf{$w_{CstAss}^+$}&\textbf{$w_{Source}^+$} & \textbf{$w_{CstDis}^+$}&\textbf{$w_{Sink}^+$}&\textbf{$w_{Moll}^+$}\\
 	=0&=1&$=2.7\times10^{-8}$& $=0$&=1& =1&=0& $=2.72\times10^{-8}$&=0&=1 \\\hline
 \end{tabular}
 \begin{flushleft} $w_*^+$ denotes the sum of the Akaike weights of scenarios including the type $*$ of the process. Type of transport - No  drift for the insoluble pool (No drift) vs Drift with almost constant speed (Cst drift) vs  Drift with variable speed (Var. drift). Type of turnover terms - Linear vs Nonlinear.  Type of assembly - Non-localized (constant) vs Localized of type Sources. Type of disassembly - Non-localized (constant) vs Localized of type Sinks vs Localized of type Mollify.   
 \end{flushleft}
 \label{Table:Importance}
 \end{adjustwidth}
 \end{table}
\begin{table}[!h]
 \begin{adjustwidth}{-1.25in}{0in}
 	\caption{{\bf Importance of the type of process.}}
 	\begin{footnotesize}
 	\begin{tabular}{|l|l|l|l|l|l|l|}
 	\hline
 	\textbf{Scenario $i$}&\textbf{Cst Ass. Dis.}&\textbf{Cst Ass. Sinks}&\textbf{Cst Ass. Moll.}&\textbf{Sources Cst Dis.}&\textbf{Sources Sinks}&\textbf{Sources Moll.}\\\hline
 	1&1&&&&&\\ 
 	2&&1&&&&\\ 
 	3&&&1&&&\\ 
 	4&&&&1&&\\ 
 	5&&&&&1&\\ 
 	6&&&&&&1\\ 
 	7&1&&&&&\\ 
 	8&&1&&&&\\ 
 	9&&&1&&&\\ 
 	10&&&&1&&\\ 
 	11&&&&&1&\\ 
 	12&&&&&&1\\ 
 	13&1&&&&&\\ 
 	14&&1&&&&\\ 
 	15&&&1&&&\\ 
 	16&&&&1&&\\ 
 	17&&&&&1&\\ 
 	18&&&&&&1\\ 
 	19&1&&&&&\\ 
 	20&&1&&&&\\ 
 	21&&&1&&&\\ 
 	22&&&&1&&\\ 
 	23&&&&&1&\\ 
 	24&&&&&&1\\ 
 	25&1&&&&&\\ 
 	26&&1&&&&\\ 
 	27&&&1&&&\\ 
 	28&&&&1&&\\ 
 	29&&&&&1&\\ 
 	30&&&&&&1\\ 
 	31&1&&&&&\\ 
 	32&&1&&&&\\ 
 	33&&&1&&&\\ 
 	34&&&&1&&\\ 
 	35&&&&&1&\\ 
 	36&&&&&&1\\\hline
 	\hline
 	&\textbf{$w_{CstAssDis}^+$}&\textbf{$w_{CstAssSink}^+$}&\textbf{$w_{CstAssMoll}^+$}&\textbf{$w_{SourceDis}^+$}&\textbf{$w_{SourceSink}^+$}&\textbf{$w_{SourceMoll}^+$}\\
 	&$=2.7\times10^{-8}$&=0&=1&=0&=0&=0\\\hline
 	\end{tabular} \end{footnotesize}
 	\begin{flushleft}
 	 Combinations of assembly/disassembly rate profiles  - Constant assembly and disassembly rates vs constant assembly rate and disassembly rate of type Sinks vs constant assembly rate and disassembly rate of type Mollify vs assembly rate of type Sources and constant disassembly rate vs assembly rate of type Sources and disassembly rate of type Sinks vs assembly rate of type Sources and disassembly rate of type Mollify.   Top: The numerical value 1 denotes that the combination of processes of interest is in Scenario $i$. Bottom: $w_*^+$ denotes the sum of Akaike weights of scenarios including the process combination as indicated at the top.
 	\end{flushleft} 
 	\label{Table:AssemblyDisassembly}
 	\end{adjustwidth}
 \end{table}
\newpage 
 
\begin{itemize}
\item {\bf What is the most likely type of transport for the keratin material in cells given the experimental data and the model collection considered?}  {\it Is it that there is no drift of the insoluble pool (diffusion only for both pools), drift with an almost constant speed for the insoluble pool or drift with a variable speed for the insoluble pool}?  Each category includes 12 scenarios (see $3^{rd}$ to $5^{th}$ columns in Fig.~\ref{fig:Hypotheses}). From the Akaike weights, it is obvious that scenarios including diffusion only (with no drift) are not supported by experimental data since $w_{NoDrift}^+=0$ (Table~\ref{Table:Importance}). Diffusion alone is not enough to explain the experimental data. Given the data and the model collection, the type of transport can be ordered as follows (Table~\ref{Table:Importance}):
\begin{center}
	{\it Drift with almost constant speed $>$ Drift with variable speed $\gg$ No  drift}.
\end{center}

\item {\bf What is the most likely type of turnover between the soluble and insoluble pools given the experimental data and the model collection considered?} {\it Is linear exchange or Michaelis-Menten type (underlying an enzyme activity) more likely}? Each category includes 18 scenarios (see $6^{th}$ to $7^{th}$ columns in Fig.~\ref{fig:Hypotheses}). From Table~\ref{Table:Importance}, enzymatic activities are more likely to be present. Hence
\begin{center}
	{\it Nonlinear exchange $\gg$ Linear exchange.}
\end{center}

\item {\bf What is the most likely rate profile of the assembly process of the keratin material in cells given the experimental data and the model collection considered?} {\it Is a constant assembly rate or assembly mainly localized at the cell membrane periphery more likely}? Each category includes 18 scenarios (see $8^{th}$ to $9^{th}$ columns in Fig.~\ref{fig:Hypotheses}).  From Table~\ref{Table:Importance}, the non-localized rate of assembly is preferred to the rate of the localized assembly; $w_{CstAss}^+>w_{Source}^+$. Hence
\begin{center}
	{\it Non-localized assembly $\gg$ Assembly localized at cell membrane periphery.}
\end{center}

\item  {\bf What is the most likely rate profile of the disassembly process of the keratin material in cells given the experimental data and the model collection considered?} {\it Is a constant disassembly, disassembly of type Sinks or disassembly localized around the nucleus more likely}? Each category is composed of 12 scenarios (see $10^{th}$ to $12^{th}$ columns in Fig.~\ref{fig:Hypotheses}). From Table~\ref{Table:Importance}, none of the  scenarios that include the disassembly rate of type Sinks is supported by experimental data.
\begin{center}
	{\it Disassembly localized around the nucleus $>$ Non-localized disassembly $\gg$ Sinks-type profile.}
\end{center}

\item {\bf What is the most likely combination of the assembly and disassembly rate profiles given the experimental data and the model collection considered?}  {\it Is a combination of constant assembly and disassembly rates, a constant assembly rate and disassembly rate of type Sinks, a constant assembly rate and disassembly rate of type Mollify, an assembly rate of type Sources and constant disassembly rate, an assembly rate of type Sources and disassembly rate of type Sinks or an assembly rate of type Sources and disassembly rate of type Mollify more likely?} Now, interactions between the assembly and disassembly processes are investigated. Six categories are defined. In each category, there are 6 scenarios as listed in the top of Table~\ref{Table:AssemblyDisassembly}. Overall, as in Scenario 21, the combination of non-localized assembly and disassembly localized around the nucleus is preferred (Table~\ref{Table:AssemblyDisassembly}).
\begin{center}
{\it Non-localized assembly and disassembly localized around the nucleus	$>$ Non-localized assembly and disassembly $\gg$ Any other combination.}
\end{center}
\end{itemize}

\section*{Discussion}
The primary aim of the present work is to investigate, through mathematical modeling driven by experimental observations, mechanisms contributing to the organization of the keratin material in epithelial cells and subsequently to test the biological model for the keratin dynamics proposed by Leube and Windoffer {\it et al.} in 2011 \cite{Leube2011,Windoffer2011}.  From first principles, we formulate a collection of mathematical models capturing various combinations of biological processes describing the spatio-temporal dynamics of the keratin material in epithelial cells. By using techniques in parameter estimation, we find optimal reaction kinetic parameter values for each model that best captures experimental observations. We go one step further and employ an information-theoretic approach for model selection to determine the best model among all the best fit models that captures the key processes of the experimental data.

 As previously highlighted, the model framework and the description of processes considered are driven by experimental data and observations.  Experimental data used in this study provide the spatial distribution of the assembled keratin concentration at 24 and 48 hours. As only macroscopic information on the keratin organization is available, the models only describe the keratin material as soluble or assembled and consider exchanges between these soluble and insoluble pools combined with transport events. For instance, as experimental data identify the existence of regions with preferential assembly and disassembly (Fig.~\ref{Figure_Data}.3), the assembly and/or disassembly processes are assumed to be localized in some scenarios. When observing the perpetual inward motion of the keratin network (see \nameref{SI0}), the ``natural assumption'' for the transport of the assembled keratin would be the existence of an inward active transport. However, we decide to consider scenarios with only passive transport (no active transport). Why?  First, it is well known that combining appropriate reaction terms and diffusion (only) can lead to the emergence of complex behavior such as traveling waves and/or pattern formation (see some examples in \cite{Anotida2003}). Secondly, to reinforce the predictive power of our study. In our approach, we do not only design mathematical models but also calibrate models (parameter estimation) and then evaluate (model selection) how each model performs. To highlight and confirm the importance or existence of a given process, scenarios in which the process is absent must  also be considered and evaluated. Our strategy of considering ``unrealistic scenarios'' and ``realistic scenarios'' in the collection of models to be evaluated mimics the biological experimental protocols such as knockout and controls. For instance, it has been shown that the keratin assembly / disassembly depend on post-translational modifications of keratins due to enzymatic activities \cite{Inagaki2006}. In our study, a total of 36 scenarios that divide into 18 scenarios with a turnover of type linear (no enzymatic activities) and 18 scenarios with a turnover of type nonlinear (enzymatic activities) are investigated. There is a one-to-one correspondence between the 18 scenarios of the 2 groups. When model performances (Akaike weights) are compared, scenarios having a nonlinear turnover perform better than the corresponding ones with a linear turnover. Hence, the enzymatic activity is detected by the model selection as existing but also preferable. This outcome allows us to judge that our approach combining the mathematical modeling and model selection performs correctly, it gives us the confidence and trust in our conclusions.

 Following this methodology, it turns out that Scenario 21 was the best of the best fits. Scenario 21 is in good agreement with the biological model proposed in \cite{Leube2011,Windoffer2011}. Scenario 21 and the biological model share the following common key features: {\it diffusion of the soluble pool,  inward active transport of the insoluble pool and disassembly of the insoluble pool localized around the nucleus}. Moreover, as experimentally observed, Scenario 21 preserves the repartition of the keratin material over time; the keratin material is mainly insoluble in epithelial cells. Both the proposed biological model and Scenario 21 hypothesize an inward active transport for the assembled keratin material. Examining the Akaike weights of the models with no active transport (Scenarios 1 to 12 in Table~\ref{Table:Akaike}, $1^{st}$ and $2^{nd}$ columns in Fig.~\ref{fig:NonL} and $1^{st}$ to $3^{rd}$ columns in Table~\ref{Table:Importance}), we go one step further and show that the active transport is a requirement to explain the experimental data  and that the experimentally observed inward motion of the assembled keratin is not an emergent phenomenon but it is due to an active transport. Furthermore, comparing the characteristic time scales of processes, we establish that the keratin dynamics are mainly controlled by the active transport in Scenario 21. Interestingly, Scenario 21 concurs with the proposed biological model about the existence of a disassembly localized in the perinuclear region. An experimental protocol must now be developed to further work out the details of  this conclusion. It is worth pointing out that some characteristics of the proposed biological model were not tested due to the form of the mathematical models. For instance, the scenarios considered here describe the turnover in terms of the assembly of soluble proteins and disassembly of aggregated proteins. In the biological model, the nucleation of filaments, {\it i.e.} the initiation of filaments, is explicitly described and localized at the cell periphery. However, the nucleation process is not explicitly described in any of the present models. It is worth mentioning that at the beginning of this work a nucleation term was included in the models; the effect of this term did not change significantly the results; hence, to reduce the complexity of models and the number of parameters the nucleation term was subsequently dropped. Investigation, by mathematical modeling, of the nucleation process is currently being carried out in a separate and on-going study.

 When only scenarios with a variable drift are considered (Scenarios 25 to 36, $5^{th}$ and $6^{th}$ columns in Fig.~\ref{fig:NonL}), the preferred combination of assembly and disassembly rate profiles is a constant rate of assembly and disassembly as in Scenario 31, which is the second best model. 
The best model, Scenario 21, and the second best model, Scenario 31, share the non-compartmentalization of the assembly process. However, in Scenario 21 the  active transport has an almost constant speed whereas in Scenario 31 the speed is variable. For the disassembly, this process is non-localized in Scenario 31 whereas it is localized at the perinuclear region in Scenario 21. The spatial variability of the speed $u(x)$ in Scenario 31, in particular, the almost-zero speed at the nucleus locations (see for $x\in [-7.5,7.5]$ in Fig.~\ref{Figure_AllSpeed}) ``compensates''  for the non-compartmentalization of the disassembly process. After comparing the features of  Scenarios 21 and 31, we inspect their profiles. The profile obtained with Scenario 21 fits very well the experimental data on non-perinuclear locations (see for $x\notin[-7.5,7.5]$ in Fig.~\ref{fig:NonL}.33). On the other hand, the profile resulting from Scenario 31 fits very well experimental data on perinuclear locations (see for $x\in[-7.5,7.5]$ in Fig.~\ref{fig:NonL}.23). As an improvement of Scenario 21, we would expect that a wider decay in speeds around the nucleus modeled, for instance, with a smaller value of $a$ in \eqref{eq:ConstantSpeed} would result in a better fit of the perinuclear region.

Scenarios 29 and 35 include the variable speed measured from experimental data, a localized assembly rate having the spatial profile deduced from Sources and a localized disassembly rate whose  shape is derived from the profile of the Sinks. All the characteristics extracted from the experimental data are included in Scenario 29 (linear model) and Scenario 35 (nonlinear model). If one wanted the best model that takes into account all the experimental features, then Scenario 29 or 35 would be the best scenario. Scenario 29 is ranked $27^{th}$ and Scenario 35 is ranked $6^{th}$ (Table~\ref{Table:Akaike}). Similarly to the general trend, for the same set of assumptions, the use of the Michaelis-Menten type turnover term generally provides a better representation of experimental data than the use of the linear turnover term. The evidence ratio of Scenarios 35 and 29 $\omega_{35}/\omega_{29}$ is ridiculously large; Scenario 35 is much more adequate to represent the experimental data than Scenario 29. However, Scenario 35 is still only ranked $6^{th}$.  The failure/mismatch of Scenario 35 might be explained by the redundancy of the information existing in the variable speed and the net assembly/disassembly region profiles extracted from the experimental data. Furthermore, according to the Akaike weights, disassembly rates of type Sinks (the type deduced from experimental data) are less likely to occur given the experimental data and the collection of models considered. A similar conclusion is obtained for the assembly rate of type Sources. This could be a consequence of our too conservative interpretation of the regions of preferential assembly or disassembly. In our assumptions, in zones of preferential assembly, the disassembly rate is set to be about null and vice versa (Fig.~\ref{Figure_Data}.3).

 In summary, to model the keratin dynamics in epithelial cells we characterized the keratin material into two pools, the soluble and the insoluble pool; and, the events considered are the turnover and transport for both pools. The modeling assumptions used, for instance, the diffusion of the soluble pool, are based on biological observations and experimental data. The collection of the models considered in this study is designed to answer a set of questions such as ``what is the mode of transport of the keratin material in epithelial cells?''. After optimizing parameter values, model selection and evaluation methods applicable to non-nested models are used to discriminate between the candidate models, identify the best model and quantify how models under consideration are adequate to explain the experimental data. Note that the ranking of the models (scenarios) and the relative importance of the different types of processes (Tables \ref{Table:Akaike}-\ref{Table:AssemblyDisassembly}) are only valid in the context of the experimental data and the set of the candidate models considered here. For instance, considering other hypotheses such as the non-negligence of the anterograde motion of the assembled keratin material or the stabilization (protection against disassembly) of the keratin filaments involved in the nuclear cage would have led to a different collection of scenarios in which our best scenario could have failed to be the best one. Furthermore, as some modeling assumptions are directly derived from experimental data, missing information in experimental data might have been prejudicial for the correct approximation / modeling of processes; for instance, missing information about the speed of assembled keratin in the cell periphery result in small values for the variable speed close to the plasma membrane (Fig.~\ref{Figure_Data}.2). Keeping in mind the limitations of our approach, important conclusions have been reached such as
\begin{itemize}
\item an active transport of the assembled keratin material is required, thereby confirming recent experimental observations \cite{Leube2011,Windoffer2011},
\item enzymatic activities regulating the assembly / disassembly are more likely to occur,
\item the assembly process is more likely to be non-compartmentalized in cells,
\item   last but not least, a unique best model strongly supported by experimental data is identified, this scenario is in good agreement with the biological model previously proposed in \cite{Leube2011,Windoffer2011}. Interestingly, the best scenario supports the perinuclear localization of the disassembly process hypothesized in the biological model.
\end{itemize}

Mathematical models of the keratin intermediate filament organization in cells were previously proposed \cite{Portet03b,Portet2008b,PortetArino2009,Sun2011,Kim2010}; however, none of these models included the effects of transport  of the assembled material on its organization. More importantly, only the behavior of models were validated qualitatively, no comparisons to experimental data were carried out. It is worth noting that other studies of the intermediate filaments dynamics combining mathematical modeling and experimental data approaches were carried out but on neurofilaments in neurons (see for example \cite{Brown2005,Li2014}). 

\section*{Conclusion}
Given the experimental data published in \cite{Moch2013}, through modeling and simulations, we investigate the effects of the interplay between turnover and transport on the keratin spatio-temporal organization  in epithelial cells. Out of all the scenarios investigated, a scenario strongly supported by experimental data is found that best captures most of the hallmarks of the  experimental observations. This scenario predicts the diffusion of soluble keratin, an  inward active transport of the assembled keratin  and the disassembly localized around the nucleus triggered by enzymatic activities as well as the assembly process that is non-compartmentalized over the cell. The value of our models is reflected in their predictive nature; first, the approach predicts the localized disassembly at the perinuclear region and second, that the experimentally observed inward motion is not an emergent behavior but that it is an inherent property of the organization of the keratin material in epithelial cells and it is due to an active transport.

\subsection*{Appendix 1}\label{SI1}
Based on experimental observations, the speed of the assembled keratin material is assumed to decay to almost zero around the nucleus. The nuclear envelope positions are at $x=\pm 7.5 \mu m$. In order to describe the decay of the speed around the nucleus such that it is almost constant, the following function is used:
\begin{align}v(x)=u(1-exp(-a x ^2)),
\label{eq:ConstantSpeed}
\end{align}
with $a=0.05$ and $u$  being in the range given in Table~\ref{tab:ModelParameter}. For numerical simulations, $u$ is set to $0.0025\mu m/ s$  which represents the average value of speeds measured in \cite{Moch2013}.

\begin{figure}[!h]
	\includegraphics[scale=.5]{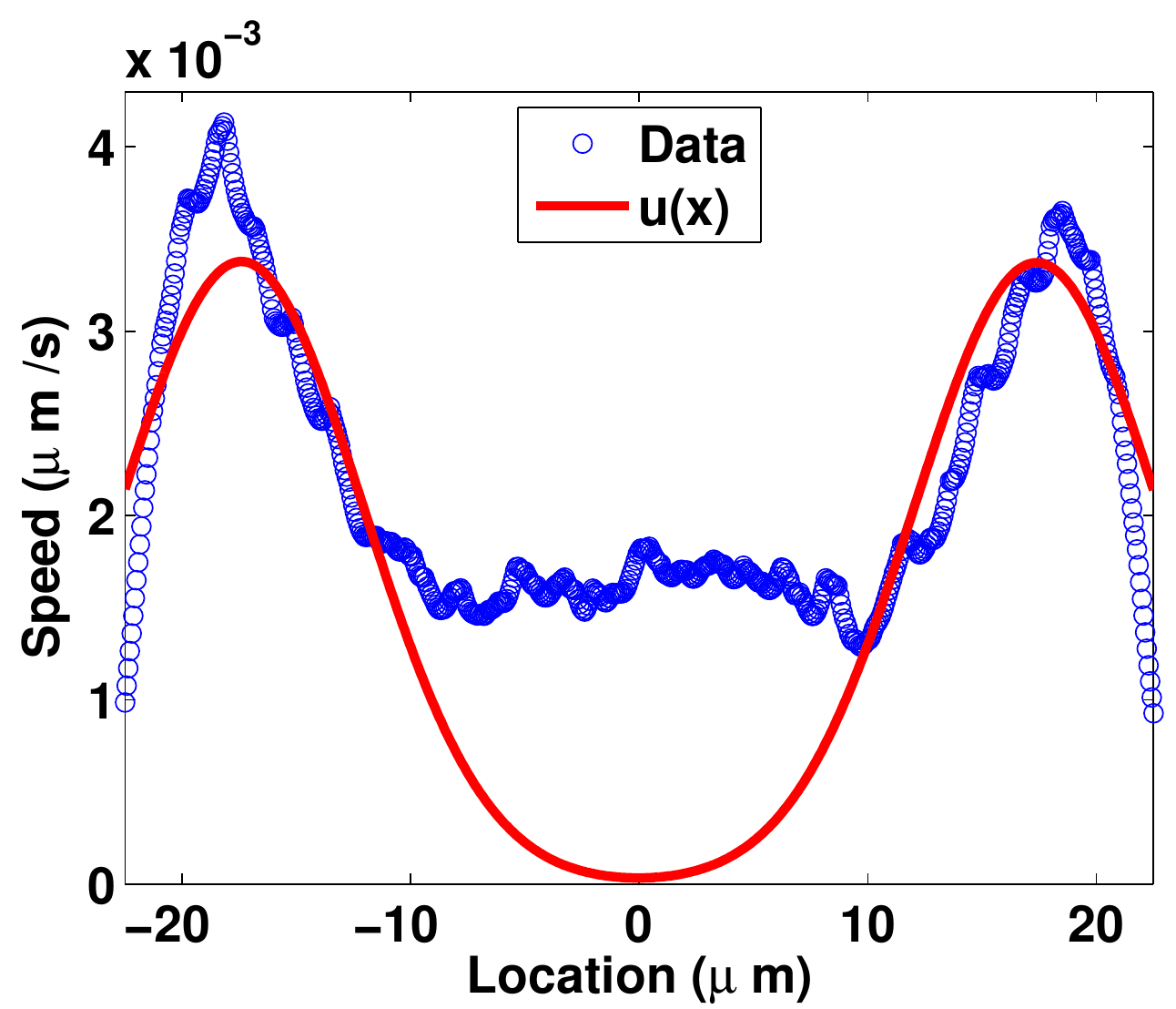}
	\caption{{\bf Estimate of the space-dependent speed.} Estimate of the space-dependent speed given in \eqref{eq:estimatev(x)WholeCellSymmetrical} for the variable speed $u(x)$ of the active transport of the insoluble pool.}
	\label{fig:SpeedEstimated_Portet}
\end{figure}

For the variable speed case, a symmetrical function over the spatial domain $\Omega$ is sought to describe the speed $u(x)$.  Averaging over the symmetrical spatial locations values of the average speed measured in \cite{Moch2013} (Fig.~\ref{Figure_Data}.2) and curve fitting with a sum of Gaussian functions of  these values, an estimate of $u(x)$ is obtained and expressed as follows:
\begin{align}
u(x) = & \sum _{i=1}^2 a_i\times exp(-((x-b_i)/c_i)^2), \label{eq:estimatev(x)WholeCellSymmetrical}
\end{align}
with the following coefficients: $a_1 =    0.003372$, $b_1 =17.39$,  $c_1 = 7.577$, $a_2 =  0.003378$, $b_2 =  -17.41$ and $c_2 = 7.546 $. 
The estimate $u(x)$ is almost zero at the nucleus-locations and is symmetrical  around the center of the cell (Fig.~\ref{fig:SpeedEstimated_Portet}).

For numerical simulations, when needed, the following function
$$s(x)=\frac{2}{1+e^{-x/a}}-1$$
with $a=1$ is used as a ``smooth analogue'' of  $sgn(x)$.

\subsection*{Appendix 2}\label{SI2}
\begin{figure}[!h]
\includegraphics[scale=.5]{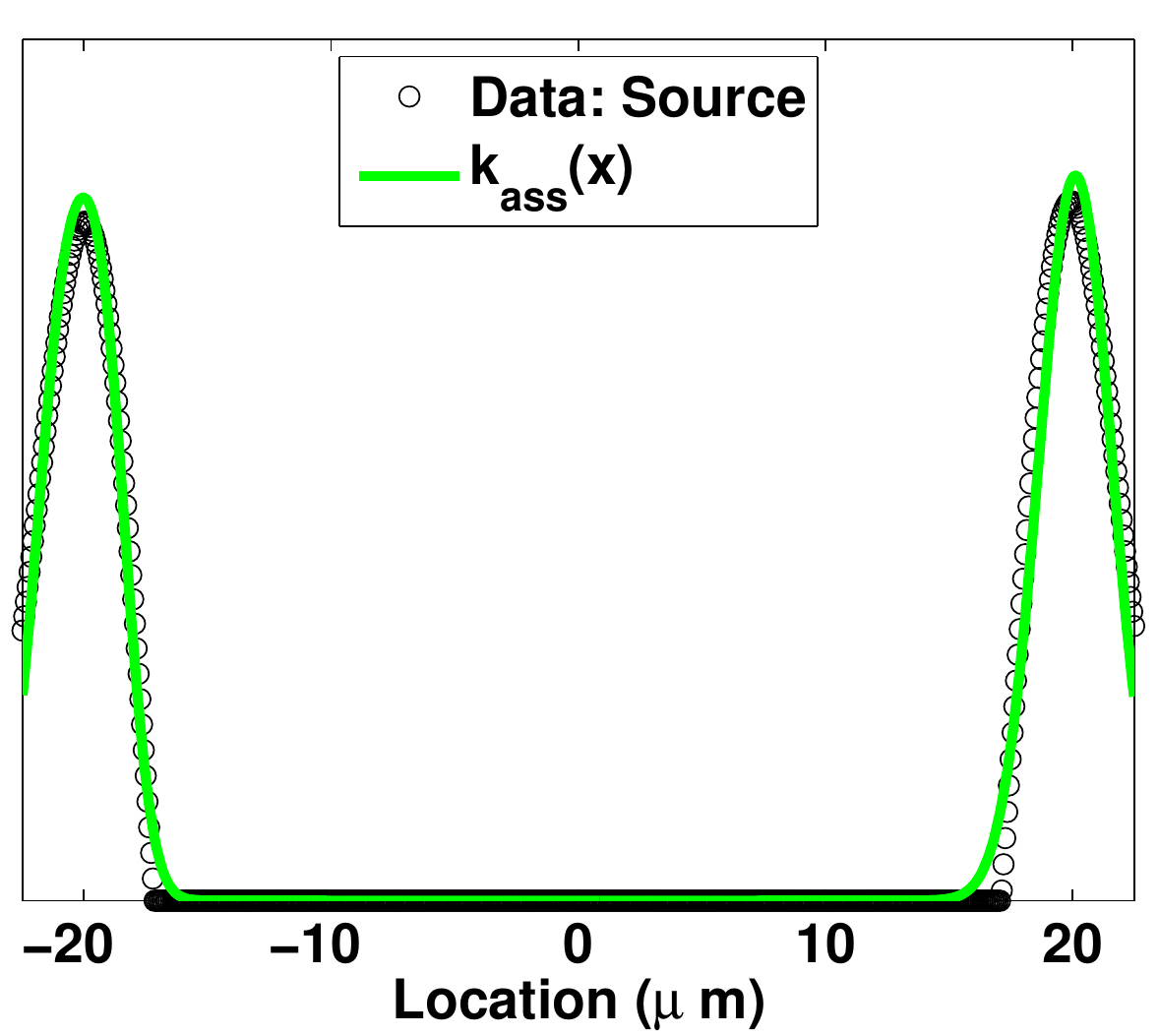}
\caption{{\bf Profile for localized assembly rate of type Sources.} The profile for $k_{ass}(x)$ defined in \eqref{eq:SourceKAss} with $k_{max}=1$, obtained by fitting the profile of assembly regions Sources published in \cite{Moch2013} and shown in Fig.~\ref{Figure_Data}.3.}
\label{fig:SourceKAss}
\end{figure}

To obtain the space-dependent function $k_{ass}(x)$ the profile of the regions of assembly (Sources) published in \cite{Moch2013} (Fig.~\ref{Figure_Data}.3) is first made symmetrical by averaging values at the symmetrical spatial locations and then the symmetrical Sources profile is fitted using  the sum of Gaussian functions:
\begin{align}
k_{ass}(x) = &k_{max} \sum _{i=1}^3 a_i\times exp(-((x-b_i)/c_i)^2) +k_{baseline}.\label{eq:SourceKAss}
\end{align}
The best fit is obtained with the following coefficients: $a_1 = 1.684\times 10^{5}$, $b_1 =20.13$, $c_1 =2.105$, $a_2 = 8.493\times 10^{4}$, $b_2 = -19.04$, $c_2 = 1.374$, $a_3 =   1.333\times 10^{5}$, $b_3 = -20.74$, and $c_3 = 1.734$. The value of the coefficient $k_{max}$ determines the maximal value of the peaks. When  $k_{max}=\frac{2L(k_{ass}-k_{baseline})}{\sqrt{\pi}\sum_{i=1}^3a_i|c_i|}$ with $k_{baseline}=k_{ass}\times 10^{-2}$, the total amount of assembly over the cell is the same  as  that of the case of a constant assembly of level $k_{ass}$. The value $k_{ass}$ in $k_{max}$ is determined by fitting model solutions to experimental data. An illustration of the shape of $k_{ass}(x)$ is given in Fig.~\ref{fig:SourceKAss}.

\subsection*{Appendix 3}\label{SI3}
\begin{figure}[!h]
\includegraphics[scale=.5]{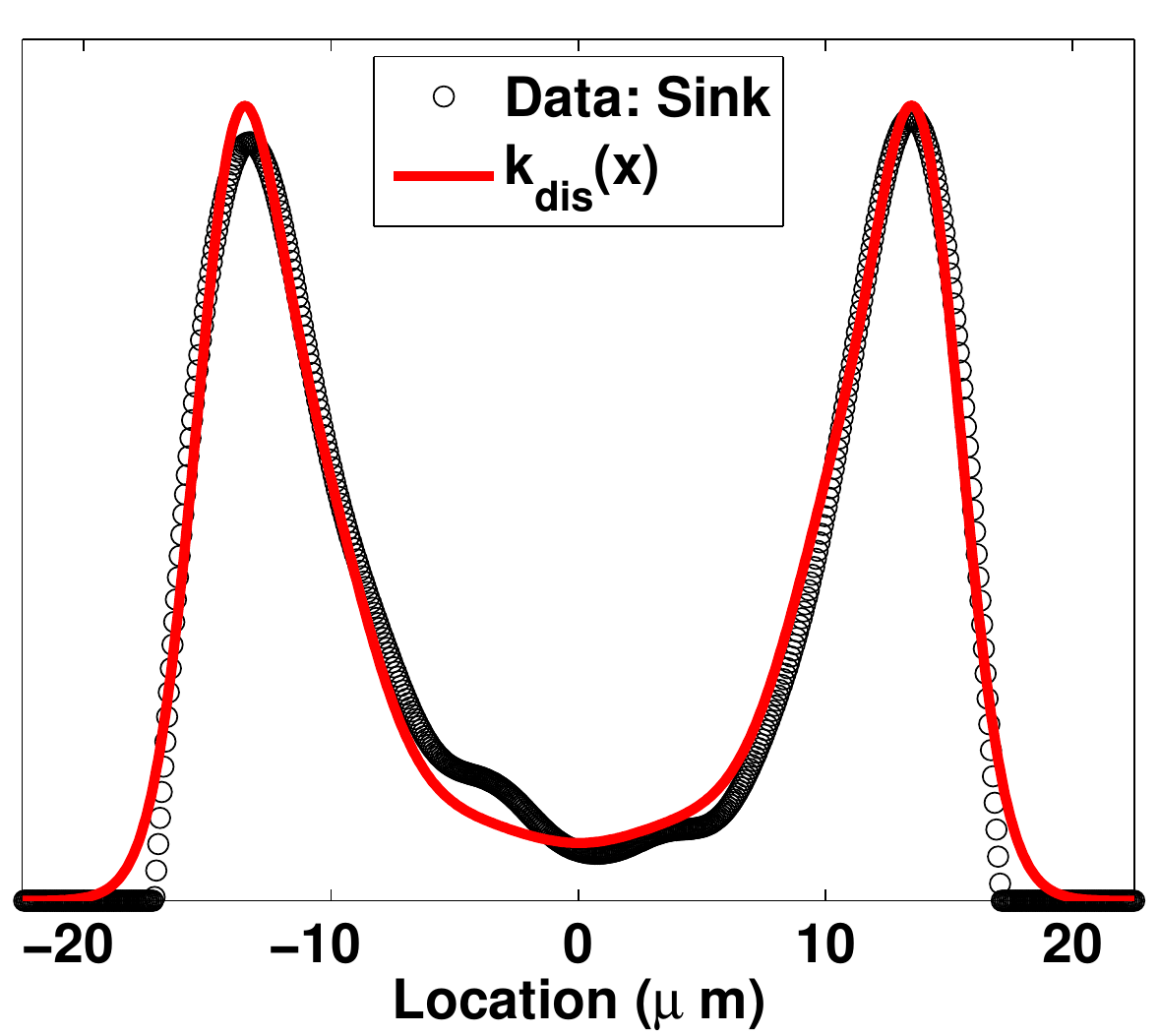}
\caption{{\bf Profile for localized disassembly rate of type Sinks.} The profile for $k_{dis}(x)$ defined in \eqref{eq:SinkKDis} and $k_{max}=1$, obtained by fitting the profile of disassembly regions Sinks published in \cite{Moch2013} and shown in Fig.~\ref{Figure_Data}.3.}
\label{fig:SinkKDis}
\end{figure}

To obtain the first space-dependent function $k_{dis}(x)$, the profile of regions of disassembly (Sinks) published in \cite{Moch2013} (Fig.~\ref{Figure_Data}.3) is made symmetrical and fitted using  the  sum of Gaussian functions:
\begin{align}
k_{dis}(x) = &k_{max} \sum _{i=1}^8 a_i\times exp(-((x-b_i)/c_i)^2). \label{eq:SinkKDis}
\end{align}
The best fit of the symmetrical profile of Sinks is obtained with the following coefficients:
$a_1 =  -3.383\times 10^{5}$, $b_1 =-11.51$, $c_1 =4.158$, $a_2 =-1.004\times 10^{5}$, $b_2 =-14.03$, $c_2 =2.236$, $a_3 =2.445\times 10^{5}$, $b_3 =-11.39$, $c_3 =4.339$, $a_4 =  -2.032\times 10^{4}$, $b_4 = 5.922$, $c_4 =5.385$, $a_5 =  -2.058\times 10^{4}$, $b_5 =-6.077$, $c_5 = 5.778$, $a_6 =3.793\times 10^{5}$, $b_6 =11.37$, $c_6 =4.322$, $a_7 = -1.008\times 10^{5}$, $b_7 =14.03$, $c_7 =2.238$, $a_8 = -4.742\times 10^{5}$, $b_8 = 11.46$, and  $c_8 =4.196 $. The coefficient $k_{max}$ determines the amplitude of the peaks. When  $k_{max}=\frac{2Lk_{dis}}{\sqrt{\pi}\sum_{i=1}^8a_i|c_i|}$, the total amount of disassembly over the cell is the same as  that of  the case of the constant function of level $k_{dis}$. The value $k_{dis}$ in $k_{max}$ is determined by fitting model solutions to experimental data. An illustration of the shape of $k_{dis}(x)$ obtained from Sinks is given in Fig.~\ref{fig:SinkKDis}.

A second function is hypothesized to represent the localized disassembly $k_{dis}(x)$ around the nucleus. A mollified piecewise function is used and is of the form
\begin{align}
k_{dis}(x)=\left \{
\begin{array}{lc}
k_{baseline}, &  x <a_1 - \epsilon,\\
-\frac{k_{max}}{4\epsilon(a_1-a_2)}x^2 -\frac{k_{max}(\epsilon-a_1)}{2\epsilon(a_1-a_2)} x &\\
\quad +  \frac{-(\epsilon - a_1)^2k_{max}+4\epsilon(a_1-a_2)k_{baseline}}{4\epsilon(a_1-a_2)},&a_1 - \epsilon\leq x <a_1 + \epsilon,\\
k_{baseline}+k_{max} \frac{x-a_1}{a_2-a_1}, & a_1 + \epsilon\leq x<a_2 - \epsilon,\\
-\frac{k_{max}}{4\epsilon(a_2-a_1)}x^2 +\frac{k_{max}(\epsilon+a_2)}{2\epsilon(a_2-a_1)} x &\\
\quad  -\frac{((\epsilon - a_2)^2+4\epsilon a_1)k_{max}+4\epsilon(a_1-a_2)k_{baseline}}{4\epsilon(a_2-a_1)},&a_2 - \epsilon\leq x <a_2 + \epsilon,\\
k_{baseline}+k_{max}, & a_2 + \epsilon \leq x< a_3 - \epsilon, \\
\frac{k_{max}}{4\epsilon(a_3-a_4)}x^2 +\frac{k_{max}(\epsilon-a_3)}{2\epsilon(a_3-a_4)} x &\\
\quad  +\frac{((\epsilon + a_3)^2-4\epsilon a_4)k_{max}+4\epsilon(a_3-a_4)k_{baseline}}{4\epsilon(a_3-a_4)},&a_3 - \epsilon\leq x <a_3 + \epsilon,\\
k_{baseline}+k_{max}-k_{max} \frac{x-a_3}{a_4-a_3}, & a_3 + \epsilon \leq x< a_4 -\epsilon, \\
\frac{k_{max}}{4\epsilon(a_4-a_3)}x^2 -\frac{k_{max}(\epsilon+a_4)}{2\epsilon(a_4-a_3)} x &\\
\quad +  \frac{(\epsilon + a_4)^2k_{max}+4\epsilon(a_4-a_3)k_{baseline}}{4\epsilon(a_4-a_3)},&a_4 - \epsilon\leq x <a_4 + \epsilon,\\
k_{baseline}, & a_4 +\epsilon \geq x ,
\end{array}
\right .
\label{eq:LocDisWhole}
\end{align}
where $a_1=-15 \mu m$, $a_2= -7.5 \mu m$, $a_3=7.5 \mu m$, $a_4=15\mu m$ and $\epsilon = 1$. When $k_{max}=2(k_{dis}-k_{baseline})$ with $k_{baseline}=k_{dis}\times10^{-2}$, the total amount of disassembly over the cell is the same  as  that of the case of a constant disassembly of level $k_{dis}$; $k_{dis}$ is determined by fitting the model solutions to experimental data. An illustration of the shape of \eqref{eq:LocDisWhole} 
 is given in Fig.~\ref{fig:RateProfile}.

\subsection*{Appendix 4}\label{SI4}
To define the initial condition $f_0(x)$, the experimental profile \cite{Moch2013} of the assembled keratin material measured at 24 hours is used (gray circles in Fig.~\ref{Figure_Data}.1). The polynomial $P(x) = \sum_{i=0}^{8}p_ix^{i}$ for which $p_8 =5.441\times 10^{-8}$, $p_7 =3.397\times 10^{-21}$, $p_6 = -5.379\times 10^{-5}$, $p_5 =-2.077\times 10^{-18}$, $p_4 =0.01062$, $p_3 =2.801\times 10^{-16}$ $p_2 =0.4104$, $p_1 =1.4\times 10^{-14}$ and $p_0 =506.5$ fits well the data at 24 hours (gray curve in Fig.~\ref{Figure_Data}.1). The function $P(x)$ is then modified to obtain a function satisfying the boundary conditions defined in \eqref{eq:BC}. 
The initial condition $f_0(x)$ describing the profile of the insoluble pool at $24$ hours is expressed as follows:
\begin{align}
f_0(x)=\left \{
\begin{array}{lc}
c, & x<a_1 - \epsilon, \\
\frac{P(a_1+\epsilon)-c}{4\epsilon^2}x^2+\frac{(\epsilon-a_1)(P(a_1+\epsilon)-c)}{2\epsilon^2}x &\\ \quad +\frac{(\epsilon-a_1)^2P(a_1+\epsilon)+(3\epsilon^2+2a_1\epsilon-a_1^2)c}{4\epsilon^2}, &  a_1 - \epsilon\leq x<a_1 + \epsilon, \\
P(x), & a_1 + \epsilon\leq x<a_2 - \epsilon, \\
 -\frac{P(a_2-\epsilon)-c+2\epsilon D(a_2-\epsilon)}{4\epsilon^2}x^2 &\\
 -\frac{(\epsilon-a_2)(P(a_2-\epsilon)-c)-2\epsilon a_2D(a_2-\epsilon)}{2\epsilon^2}x &\\
  \quad + \frac{(3\epsilon^2+2a_2\epsilon-a_2^2)P(a_2-\epsilon)}{4\epsilon^2} & \\
  \quad +\frac{2\epsilon(\epsilon^2-a_2^2) D(a_2-\epsilon)+(\epsilon-a_2)^2c}{4\epsilon^2}, & a_2 - \epsilon\leq x<a_2 + \epsilon, \\
 c, & a_2 + \epsilon\leq x,
\end{array} 
\right .
\label{eq:Initial}
\end{align} 
with $a_1=-21\mu m $, $a_2=-a_1$, $\epsilon=1$, $c=50\mu M$ and  $D(x)= \sum_{i=1}^{8}ip_ix^{i-1}$. The function $f_0(x)$ is graphed in Fig.~\ref{fig:Initial_Portet}.

The average profile of the assembled keratin material computed on  the normalized cells after 48 hours of seeding \cite{Moch2013} is represented by:
\begin{align}
f_{final}(x)=  \sum_{i=0}^{4}p_ix^{i},
\label{eq:Final}
\end{align}
where $p_4 =-0.003255$, $p_3 =2.61\times 10^{-17}$, $p_2 =0.4899$, $p_1 =1.558\times 10^{-15}$ and $p_0 =604.1$. $f_{final}(x)$ is graphed in Fig.~\ref{Figure_Data}.1 (black curve).

 \subsection*{Appendix 5}\label{SI5}
The system corresponding to Scenario 21 is now nondimensionalized to estimate the characteristic time scale of each of the processes involved. Consider new independent variables $z$ and $\tau$ and new dependent variables $\tilde{S}(z,\tau)$ and $\tilde{I}(z,\tau)$ defined such that:
$$x=Az, \quad t=B\tau, \quad S(x,t)=s\tilde{S}(z,\tau), \quad I(x,t)=i\tilde{I}(z,\tau).$$
Scenario 21 takes then the following form:
\begin{align*}
\frac{\partial \tilde{S}}{\partial \tau}=& \frac{BD_S}{A^2} \frac{\partial^2 \tilde{S}}{\partial z^2}- \frac{B}{s}\left(\frac{k_{ass} \tilde{S}}{k_S/s+\tilde{S}}-\frac{k_{dis}(Az)\tilde{I}}{k_I/i+\tilde{I}}\right),\\
\frac{\partial \tilde{I}}{\partial \tau}=& \frac{B\epsilon D_S}{A^2} \frac{\partial^2 \tilde{I}}{\partial z^2}+\frac{Bu}{A}sgn(z)f(Az)\frac{\partial \tilde{I}}{\partial z}+ \frac{B}{i}\left(\frac{k_{ass} \tilde{S}}{k_S/s+\tilde{S}}-\frac{k_{dis}(Az)\tilde{I}}{k_I/i+\tilde{I}}\right),
\end{align*}
with $f(Az)$ being the functional component of $v(x)$ \eqref{eq:ConstantSpeed} defined as $f(Az)=(1-exp(-a (Az) ^2))$ with $a=0.05$ and $k_{dis}(Az)$ being the function $k_{dis}(\cdot)$ defined in \eqref{eq:LocDisWhole} and evaluated at $Az$. Taking $A=\ell$ ($\ell=2L$), $B=\ell/u$, $s=k_S$ and $i=k_S$ then $$x=\ell z \quad \textrm{and} \quad t=\frac{\ell }{u}\tau$$ and the adimensional version of the system is
\begin{align*}
\frac{\partial \tilde{S}}{\partial \tau}=& \frac{1}{\epsilon Pe} \frac{\partial^2 \tilde{S}}{\partial z^2}- Da\left(\frac{ \tilde{S}}{1+\tilde{S}}-\frac{k_{dis}(Az)/k_{ass}\tilde{I}}{k_I/k_S+\tilde{I}}\right),\\
\frac{\partial \tilde{I}}{\partial \tau}=& \frac{1 }{Pe} \frac{\partial^2 \tilde{I}}{\partial z^2}+sgn(z)f(Az)\frac{\partial \tilde{I}}{\partial z}+ Da\left(\frac{ \tilde{S}}{1+\tilde{S}}-\frac{k_{dis}(Az)/k_{ass}\tilde{I}}{k_I/k_S+\tilde{I}}\right).
\end{align*}
The constant $Pe=\frac{\tau_{Diffusion}}{\tau_{Drift}}=\frac{\ell ^2 /D_I }{ \ell / u}=\frac{u \ell}{\epsilon D_S}$ is the P\'eclet number of the insoluble pool that compares active transport and diffusion processes. The characteristic length $\ell$ is chosen to be equal to the length of the spatial domain $\Omega$, $\ell=2L$. The Damk\"ohler number $Da=\frac{\tau_{Drift}}{\tau_{Reaction}}$ compares reaction and active transport processes. The characteristic active transport time is $\tau_{Drift}=\ell /u$ as defined in the P\'eclet number. The characteristic reaction time is $\tau_{Reaction}=1/\kappa$ for which $\kappa=\frac{k_{ass}}{k_S}$. Note that $sgn(z)f(Az)$ is about $\pm 1$. 

Recall the parameter values obtained from the parameter estimation:
$k_{ass}=9.3819\mu M/s$, $k_S=570.73\mu M$,  $k_I=976.07\mu M/s$ and $k_{max}=1.976 \mu M/s$ in \eqref{eq:LocDisWhole} that correspond to a constant disassembly rate of level $k_{dis}=0.9998\mu M/s$. Hence, from the definition of $k_{max}$, $k_{dis}(Az)$ can be approximated by $k_{dis}=0.9998$. The values of the  fixed parameters are $D_S=0.88 \mu m ^2 /s$, $D_I=\epsilon D_S$ with $\epsilon=9.5\times 10^{-4}$ and $u=0.0025\mu m/s$. The following estimates are then obtained:
\begin{itemize}
\item Adimensional parameters: $k_{dis}(Az)/k_{ass}=k_{dis}/k_{ass}=0.1$ and $k_I/k_S=1.7$;
\item Diffusion time scale: $\tau_{Diffusion}=\ell^2/D$.
\begin{itemize}
\item For the soluble pool: $\tau_{Diffusion}^S=\ell^2/D_S\approx  2.3\times 10^{3}s$,
\item For the insoluble pool: $\tau_{Diffusion}^I=\ell^2/(\epsilon D_S)\approx 2.4 \times 10^{6}s$;
\end{itemize}
\item Active transport time scale: $\tau_{Drift}=\ell/u =1.8\times 10^{4}s$ (time the fluid needs to flow through the characteristic length);
\item Reaction time scale: $k_S/k_{ass}\approx 61 s$ (time for reaction to equilibrate).
\end{itemize}

To determine what mode of transport (passive vs active) and what process (transport vs reaction) dominate the dynamics of the keratin material, the P\'eclet and Damk\"ohler numbers are used:
\begin{itemize}
\item $Pe=(u \ell)/(\epsilon D_S)\approx 135 \gg 1$. Diffusion is small compared to active transport for the insoluble pool, the transport of the assembled keratin material by drift is faster than by diffusion, active transport is the dominant mode of transport;
\item $Da=(\ell k_{ass})/(u k_S)\approx 296 \gg 1$.  Active transport time scale is greater than the reaction time scale; the overall process in controlled by  active transport.
\end{itemize}

\section*{Acknowledgments}
This work was initiated in June 2013 while SP was visiting the Institute of Molecular and Cellular Biology at RWTH Aachen University (Aachen, Germany). 
The authors acknowledge suggestions from anonymous referees and the editor for the critical review that helped improve the quality of the manuscript.

\bibliography{sportet_research}

\begin{thebibliography}{10}

\bibitem{Portet2008b}
M.~Beil, S.~L\"uck, F.~Fleischer, S.~Portet, W.~Arendt, and V.~Schmidt.
\newblock Simulating the formation of keratin filament networks by a
  piecewise-deterministic markov process.
\newblock {\em J. Theor. Biol.}, 256:518--532, 2009.

\bibitem{Brown2005}
A.~Brown, L.~Wang, and P.~Jung.
\newblock Stochastic simulation of neurofilament transport in axons: the
  ``stop-and-go'' hypothesis.
\newblock {\em Mol. Biol. Cell.}, 16:4243--4255, 2005.

\bibitem{Burnham2002}
K.~Burnham and D.~Anderson.
\newblock {\em Model Selection and Multimodel Inference. A Practical
  Information-Theoretic Approach}.
\newblock Springer Verlag, second edition, 2002.

\bibitem{Chou1993}
C.-F. Chou, C.~Riopel, L.~Rott, and B.~Omary.
\newblock A significant soluble keratin fraction in ``simple'' epithelial cells
  lack of an apparent phosphorylation and glycosylation role in keratin
  solubility.
\newblock {\em J. Cell Sci.}, 105:433--444, 1993.

\bibitem{Coulombe2013}
B.~Chung, J.~Rotty, and P.~Coulombe.
\newblock Networking galore: intermediate filaments and cell migration.
\newblock {\em Curr. Opin. Cell Biol.}, 25:600--612, 2013.

\bibitem{Coulombe2009}
P.~Coulombe, Kerrns M., and E.~Fuchs.
\newblock Epidermolysis bullosa simplex: a paradigme for disorder of tissue
  fragility.
\newblock {\em J. Clin. Inv.}, 119:1784--1793, 2009.

\bibitem{Dorsey1995}
R.~Dorsey and W.~Mayer.
\newblock Genetic algorithms for estimation problems with multiple optima,
  nondifferentiability and other irregular features.
\newblock {\em J. Bus. Econ. Stat.}, 13:53--66, 1995.

\bibitem{Feng2013}
X.~Feng, H.~Zhang, J.~Margolick, and P.~Coulombe.
\newblock Keratin intracellular concentration revisited: implications for
  keratin function in surface epithelial.
\newblock {\em J. Inv. Derm.}, 113:850--853, 2013.

\bibitem{Helfand2004}
B.T. Helfand, L.~Chang, and R.D. Goldman.
\newblock Intermediate filaments are dynamic and motile elements of cellular
  architecture.
\newblock {\em J. Cell Sci.}, 117:133--141, 2004.

\bibitem{Herberich2011}
G.~Herberich, R.~Windoffer, R.~Leube, and T.~Aach.
\newblock 3d segmentation of keratin intermediate filaments in confocal laser
  scanning microscopy.
\newblock In {\em Engineering in Medicine and Biology Society, EMBC, 2011
  Annual International Conference of the IEEE}, pages 7751--7754, Aug 2011.

\bibitem{Homberg2014}
M.~Homberg and T.~Magin.
\newblock Beyond expectations: novel insights into epidermal keratin function
  and regulation.
\newblock {\em Int. Rev. Cell Mol. Biol.}, 311:265--306, 2014.

\bibitem{Inagaki2006}
I.~Izawa and M.~Inagaki.
\newblock Regulatory mechanisms and functions of intermediate filaments: A
  study using site- and phosphorylation state-specific antibodies.
\newblock {\em Cancer Sci.}, 97:167--174, 2006.

\bibitem{Johnson2004}
J.~Johnson and K.~Omland.
\newblock Model selection in ecology and evolution.
\newblock {\em Trends Ecol. Evol.}, 19:101--108, 2004.

\bibitem{Kim2010}
J.S. Kim, C.-H. Lee, and P.~Coulombe.
\newblock Modeling the self-organization of keratin intermediate filaments.
\newblock {\em Biophys. J.}, 99:2748--2755, 2010.

\bibitem{Kolsch2009}
A.~K\"olsch, R.~Windoffer, and R.~Leube.
\newblock Actin-dependent dynamics of keratin filament precursors.
\newblock {\em Cell Motil. Cytoskel.}, 66:976--985, 2009.

\bibitem{Kolsch2010}
A.~K\"olsch, R.~Windoffer, T.~W\"urflinger, T.~Aach, and R.~Leube.
\newblock The keratin-filament cycle of assembly and disassembly.
\newblock {\em J. Cell Sci.}, 123:2266--2272, 2010.

\bibitem{Leube2011}
R.~Leube, M.~Moch, A.~Kolsch, and R.~Windoffer.
\newblock ``panta rhei'': Perpetual cycling of the keratin cytoskeleton.
\newblock {\em Bioarchitecture}, 1:39--44, 2011.

\bibitem{Li2014}
Y.~Li, A.~Brown, and P.~Jung.
\newblock Deciphering the axonal transport kinetics of neurofilaments using the
  fluorescence photoactivation pulse-escape method.
\newblock {\em Phys. Biol.}, 11:026001, 2014.

\bibitem{Anotida2003}
A.~Madzvamuse, A.J. Wathen, and P.K. Maini.
\newblock A moving grid finite element method applied to a model biological
  pattern generator.
\newblock {\em J. Comp. Phys.}, 190:478--500, 2003.

\bibitem{MATLAB:2012}
MATLAB.
\newblock {\em MATLAB and Statistics Toolbox Release 2013b}.
\newblock The MathWorks Inc., Natick, Massachusetts, 2013.

\bibitem{Moch2013}
M.~Moch, G.~Herberich, T.~Aach, R.~Leube, and R.~Windoffer.
\newblock Measuring the regulation of keratin filament network dynamics.
\newblock {\em Proc. Natl. Acad. Sci.}, 110:10664--10669, 2013.

\bibitem{Mohl2012}
C.~M\"ohl, N.~Kirchgessner, C.~Sch\"afer, B.~Hoffmann, and R.~Merkel.
\newblock Quantitative mapping of averaged focal adhesion dynamics in migrating
  cells by shape normalization.
\newblock {\em J. Cell Sci.}, 125:155--165, 2012.

\bibitem{Pan2013}
X.~Pan, R.~Hobbs, and P.~Coulombe.
\newblock The expanding significance of keratin intermediate filaments in
  normal and diseased epithelia.
\newblock {\em Curr. Opin. Cell Biol.}, 25:47--56, 2013.

\bibitem{PortetArino2009}
S.~Portet and J.~Arino.
\newblock An in vivo intermediate filament assembly model.
\newblock {\em Math. Biosc. Eng.}, 6:117--134, 2009.

\bibitem{Portet03b}
S.~Portet, O.~Arino, J.~Vassy, and D.~Schoevaert.
\newblock Organization of the cytokeratin network in an epithelial cell.
\newblock {\em J. Theor. Biol.}, 223:313--333, 2003.

\bibitem{Ramms2013}
L.~Ramms, G.~Fabris, R.~Windoffer, N.~Schwarz, R.~Springer, C.~Zhou, J.~Lazar,
  S.~Stiefel, N.~Hersch, U.~Schnakenberg, T.~Magin, R.~Leube, R.~Merkel, and
  B.~Hoffmann.
\newblock Keratins as the main component for the mechanical integrity of
  keratinocytes.
\newblock {\em Proc. Natl. Acad. Sci.}, 110:18513--18518, 2013.

\bibitem{Robert2014}
A.~Robert, H.~Herrmann, M.~Davidson, and V.~Gelfand.
\newblock Microtubule-dependent transport of vimentin filament precursors is
  regulated by actin and by the concerted action of rho- and p21-activated
  kinases.
\newblock {\em Fased J.}, 28:2879--2890, 2014.

\bibitem{Seltmann2013}
K.~Seltmann, A.~Fritsch, J.~Kas, and T.~Magin.
\newblock Keratins significantly contribute to cell stiffness and impact
  invasive behavior.
\newblock {\em Proc. Natl. Acad. Sci.}, 110:18507--18512, 2013.

\bibitem{Snider2014}
N.~Snider and B.~Omary.
\newblock Post-translational modifications of intermediate filament proteins:
  mechanisms and functions.
\newblock {\em Nat. Rev.}, 15:163--177, 2014.

\bibitem{Strnad2002}
P.~Strnad, R.~Windoffer, and R.E. Leube.
\newblock Induction of rapid and reversible cytokeratin filament network
  remodeling by inhibition of tyrosine phosphatases.
\newblock {\em J. Cell Sci.}, 115:4133--4148, 2002.

\bibitem{Sun2011}
C.~Sun, R.~Leube, R.~Windoffer, and S.~Portet.
\newblock A mathematical model for the keratin cycle of assembly and
  disassembly.
\newblock {\em IMA J. Appl. Math.}, 80:100--114, 2015.

\bibitem{Toivola2010}
D.~Toivola, P.~Strnad, A.~Habtezion, and B.~Omary.
\newblock Intermediate filaments take the heat as stress proteins.
\newblock {\em Trends Cell Biol.}, 20:79--91, 2010.

\bibitem{Windoffer2011}
R.~Windoffer, M.~Beil, T.~Magin, and R.~Leube.
\newblock Cytoskeleton in motion: the dynamics of keratin intermediate
  filaments in epithelial cells.
\newblock {\em J. Cell Biol.}, 194:669--678, 2011.

\bibitem{Windoffer1999}
R.~Windoffer and R.~Leube.
\newblock Detection of cytokeratin dynamics by time-lapse fluorescence
  microscopy in living cells.
\newblock {\em J. Cell Sci.}, 112:4521--4534, 1999.

\bibitem{Woll2005}
S.~Woll, R.~Windoffer, and R.~Leube.
\newblock Dissection of keratin dynamics: different contributions of the actin
  and microtubule systems.
\newblock {\em Eur. J. Cell Biol.}, 84:311--328, 2005.

\bibitem{Yoon2001}
K.~Yoon, M.~Yoon, R.~Moir, S.~Khuon, F.~Flitney, and R.~Goldman.
\newblock Insights into dynamic properties of keratin intermediate filaments in
  living epithelial cells.
\newblock {\em J. Cell Biol.}, 153:503--516, 2001.

\end{thebibliography}

%
%
%

\beginsupplement
\section*{Supporting Information Legends}
\subsection*{S1 Video}\label{SI0} 
 {\bf Dynamics of the keratin network in a cell.} Time-lapse fluorescence microscopy of hepatocellular carcinoma-derived PLC clone PK18-5 stably expressing fluorescent fusion protein HK18-YFP \cite{Strnad2002} depicting the dynamic properties of the keratin filaments over a time period of 15 hours.  Bar 10 $\mu m$.

\end{document}